\documentclass[twocolumn]{aastex62}
\pdfoutput=1 
\usepackage[version=4]{mhchem}
\usepackage{amsmath}
\usepackage{multirow}
\usepackage{hyperref}
\usepackage[utf8]{inputenc}
\usepackage{comment}

\newcommand{\revision}[1]{\textcolor{black}{#1}}
\newcommand{\revisiontwo}[1]{\textcolor{black}{#1}}

\graphicspath{{./}{figures/}}

\received{September 25, 2018}
\revised{Jan 9, 2019}
\accepted{Jan 9, 2019}

\shorttitle{Irradiated planets classification}
\shortauthors{Molaverdikhani et al.}

\begin{document}

\title{From cold to hot irradiated gaseous exoplanets: \revision{Toward an observation-based} classification scheme}

\correspondingauthor{Karan Molaverdikhani}
\email{Karan@mpia.de}

\affiliation{Max Planck Institute for Astronomy, Königstuhl 17, 69117 Heidelberg, Germany}

\author{Karan Molaverdikhani}
\affiliation{Max Planck Institute for Astronomy, Königstuhl 17, 69117 Heidelberg, Germany}

\author{Thomas Henning}
\affiliation{Max Planck Institute for Astronomy, Königstuhl 17, 69117 Heidelberg, Germany}

\author{Paul Molli\`ere}
\affiliation{Sterrewacht Leiden, Huygens Laboratory, Niels Bohrweg 2, 2333 CA Leiden, The Netherlands}


\begin{abstract}

A carbon-to-oxygen ratio (C/O) of around unity is believed to act as a natural separator of water- and methane-dominated spectra when characterizing exoplanet atmospheres. In this paper we quantify the C/O ratios at which this separation occurs by calculating a large self-consistent grid of \revision{cloud-free} atmospheric models \revision{in chemical equilibrium}, using the latest version of \emph{petitCODE}. \revision{Our study covers a broad range of parameter space: 400\;K$<$T\textsubscript{eff}$<$2600\;K, 2.0$<$log(g)$<$5.0, -1.0$<$[Fe/H]$<$2.0, 0.25$<$C/O$<$1.25, and stellar types from M to F.  We make the synthetic transmission and emission spectra, as well as the temperature structures publicly available.} We find that the transition C/O ratio depends on many parameters such as effective temperature, surface gravity, metallicity and spectral type of the host star, \revision{and could have values less, equal, or higher than unity}. By mapping all the transition C/O ratios we propose a “four-class” classification scheme for irradiated planets in this temperature range. \revision{We  find a parameter space} where methane always remains the cause of dominant spectral features.  \ce{CH4} detection in this region, or the lack of it, provides a diagnostic tool to identify the prevalence of cloud formation and non-equilibrium chemistry. As another diagnostic tool, we construct synthetic Spitzer IRAC color-diagrams showing two distinguishable populations of planets. \revision{Since most of the exoplanet atmospheres appear cloudy when studied in transmission, we regard this study as a starting point of how such a C/O-sensitive observation-based classification scheme should be constructed. This preparatory work will have to be refined by future cloudy and non-equilibrium modeling, to further investigate the existence and exact location of the classes, as well as the color-diagram analysis.}

\end{abstract}


\section{INTRODUCTION} \label{sec:intro}

\revision{The first classification of planets from a modern scientific point of view was proposed by Alexander von Humboldt and his colleagues in their book ``Cosmos'' \citep{humboldt_cosmos:_1852}. They categorized the solar system planetary objects into three classes of inner, central, and outer planets based on their apparent orbital configuration and noted the disparity of the inner (terrestrial) and outer (giant) planets densities. Remarkably, they also suggested different internal density distributions for these two classes; deduced from their different degree of oblateness.}

\revision{Adequate and accurate observations of these objects to examine these hypotheses did not take place until the 20$^{th}$ century. Even then the early studies on the constitution of the giant planets in the solar system were based on a few observations only. \citet{wildt_atmospheres_1934} summarized these studies and crystallized the dominant role of hydrogen in their make-up. He further developed this idea and proposed the presence of a core for these planets similar in structure to the terrestrial planets, covered by ice and a layer of solid hydrogen on top of it \citep{wildt_state_1938,wildt_constitution_1947}. However, it was not until a few years later when \citet{brown_compositions_1950} suggested the composition of Uranus and Neptune differ from Jupiter and Saturn and proposed that they are mainly composed of solid methane and ammonia. Later studies by \revisiontwo{ \citet{demarcus_constitution_1963} and \citet{zapolsky_mass-radius_1969} significantly improved our understanding of ``ice giants''\footnote{\revisiontwo{In early 70's the terminology became popular in the science fiction community, e.g. \citet{bova_many_1971}, but the earliest scientific usage of the terminology was likely by \citet{dunne_voyage_1978} in a NASA report.}} to make the first step toward the} classification of gaseous planets.}

\revision{This classification did not remain unmitigated after the detection of the first Hot-Jupiter \citep{mayor_jupiter-mass_1995} and demanded new investigations. Motivated by the apparent diversity of the solar system planetary objects, \citet{sudarsky_albedo_2000} proposed a classification of \revisiontwo{H/He dominated gaseous planets} based on their albedo and reflection spectra with five classes namely ``Jovians'' (T\textsubscript{eff}$\lesssim$150\;K), ``water cloud'' objects, ``clear'' objects, and class-IV/V ``roasters'' (i.e. close orbiting planets with T\textsubscript{eff}$\gtrsim$1500\;K). Although most of the observed exo-atmospheres appear to be cloudy, observations of their reflected light have proven to be challenging due to their faint signal and contamination by stellar noise \citep{martins_spectroscopic_2013,martins_evidence_2015,angerhausen_comprehensive_2015,kreidberg_exoplanet_2017}.}

\revision{In contrast, transmission and emission spectroscopy have found to be among the best techniques to study exoplanetary atmospheres. By considering the benefits of these techniques and extricating from the orbital configuration point of view,} \citet{fortney_unified_2008} argued that orbital period is a poor discriminator between “hot” and “very hot” Jupiters and proposed two classes of irradiated \revisiontwo{gaseous giants} by highlighting the importance of the insolation level and TiO/VO opacities for these objects. \revision{Despite the cloudy nature of exoplanets they ignored the cloud opacities due to their weak effects on the temperature structures and spectra \citep{fortney_comparative_2005}. By using cloud-free models they concluded that these two classes of} planets are somewhat analogous to the M- and L-type dwarfs and hence named them “pM Class” and “pL Class” planets according to their similarities. Ti and V in the colder objects, i.e. pL Class, are thought to be predominantly in solid condensates and neutral alkalis absorption lines were predicted to cause the dominant optical spectral features \revision{both in transmission and emission}. On the other hand, the hotter objects, i.e. pM Class, were predicted to present molecular bands of TiO, VO, \ce{H2O} and CO in emission due to their hot stratospheres (temperature inversion) and the presence of these molecules in the gas phase at photospheric pressures. They reported HD 149026b and HD 209458b as prototypical exoplanets for atmospheric thermal inversions, and classified them as pM Class planets. However, further observations and newer data reduction techniques provided evidence against an inversion in the case of HD 209458b \citep{diamond-lowe_new_2014}, and therefore the onset of pM Class is thought to begin at higher temperatures.

\revision{Along this line of thought, another class of \revisiontwo{gaseous giant} planets with T\textsubscript{eff}$>$2500\;K was recently proposed by \citet{lothringer_extremely_2018}. The chemistry of these extremely irradiated hot Jupiters is thought to be fundamentally different in comparison to the cooler planets. For instance, the presence of strong inversions due to the absorption by atomic metals, metal hydrides and continuous opacity sources such as \ce{H-}, and significant thermal dissociation of \ce{H2O}, \ce{TiO} and \ce{VO} on the dayside of these planets are predicted. As will be addressed in \hyperref[sec:Teff] {Section}\;\ref{sec:Teff}, we exclude this range of temperature (and hence this class of ultra-hot Jupiters) from our study.}

The complexity of irradiated planets classification is not limited to the effect of their insolation level. \citet{seager_dayside_2005} studied HD 209458b to place a stringent constraint on the \ce{H2O} absorption band depths. They proposed a new scenario where an atmospheric carbon-to-oxygen ratio C/O$\geq$1 explains the very low abundance of water vapor. The carbon-rich atmosphere scenario was in contrast to the solar C/O ratio of 0.55 \citep{asplund_chemical_2009} that was used in the exoplanets models prior to their study. The chemistry of such atmospheres were also predicted to be significantly different in comparison to the atmospheres with solar or sub-solar C/O ratios. Hence the spectral differences were predicted to be observable \citep{kuchner_extrasolar_2005}. However, there has not been a robust detection of a carbon-rich exoplanet yet.

\citet{madhusudhan_c/o_2012} integrated and expanded these ideas into a two-dimensional classification scheme with four classes of irradiated \revisiontwo{gaseous} planets, with the effective temperature of the planet and the C/O ratio as the two key factors on their atmospheric characteristics. Madhusudhan reported strong \ce{H2O} features for the models with a C/O ratio of 0.5. This was in contrast to the models with C/O$\geq$1 in which their spectra showed enhanced \ce{CH4} absorption at near-infrared wavelengths. Therefore, he noted that in this scheme a natural boundary between C-rich and O-rich atmospheres is plausible at C/O$=$1. He also concluded that the strength of methane spectroscopic features depends on the C/O ratio and the temperature of the observable atmosphere, \revision{but did not investigate this quantitatively and only for limited parameter space}. His conclusions can be understood from the following net reaction (e.g. \citealt{kotz_chemistry_2014,ebbing_general_2016}), assuming thermo-chemical equilibrium:

\begin{equation} \ce{CH4 + H2O <=>T[$\gtrsim 1000K$][$ \lesssim 1000K$] 3H2 +
CO \label{eq:eq1}}.  \end{equation}

For effective temperatures T\textsubscript{eff}$\gtrsim$1000\;K \hyperref[eq:eq1]{Reaction}\;\ref{eq:eq1} is in favor of CO production and thus in a C/O$<$1 atmosphere (i.e. oxygen-rich) the excess oxygen is being sequestered in the water molecules with almost no methane in the atmosphere. In the case of a C/O$>$1 atmosphere (i.e. carbon-rich) the extra carbon is bound in the methane molecules and the atmosphere is depleted of water molecules. For T\textsubscript{eff}$\lesssim$1000\;K the net \hyperref[eq:eq1]{Reaction}\;\ref{eq:eq1} is in the direction of CO depletion and therefore both \ce{H2O} and \ce{CH4} are expected to be present in the atmosphere (see e.g. \citealt{molliere_model_2015}). However, a transition from water- to methane-dominated spectrum might still occur as C/O increases, where the \ce{CH4} spectral features become stronger than the strongest \ce{H2O} features. \revision{While exoplanets, and in particular colder ones, are expected to be cloudy, \citet{madhusudhan_c/o_2012} used cloud-free atmospheric models and argued that the gas phase chemistry and corresponding spectroscopic signatures resulting from cloud-free simulations are also applicable to cloudy atmospheres \citep{madhusudhan_model_2011}.}

\revision{Following this line of thought,} \citet{molliere_model_2015} studied an extensive grid of 10,640 \revision{self-consistent cloud-free equilibrium chemistry} models investigating how stellar type,  T\textsubscript{eff}, surface gravity (log(g)), metallicity ([Fe/H]) and C/O ratio affect the emission spectra of hot
\ce{H2}-dominated exoplanets. They inspected the synthetic emission spectra and found that the water- to methane-dominated atmosphere transition occurs at C/O$\sim$$0.9$ for relatively hot planets with T\textsubscript{eff}$>$1750\;K. This is approximately consistent with a natural boundary between these two atmospheres
at C/O$\sim$1, as was predicted by \citet{madhusudhan_c/o_2012}. However, \citet{molliere_model_2015} reported a smaller value of C/O$\sim$$0.7$ for planets with an effective temperature of 1000\;K$<$T\textsubscript{eff}$<$1750\;K mainly due to oxygen being partially bound in enstatite (\ce{MgSiO3}) and other oxygen bearing condensates. They also predicted transition C/O ratios \footnote{From now on we call these C/O
ratios, ``transition" C/O ratio, (C/O)\textsubscript{tr}} as low as 0.7 for colder planets, i.e. T\textsubscript{eff}$<$1450\;K, with strong dependency on the surface gravity and atmospheric metallicity.

\revision{Given the lack of a quantitative study on the transition C/O ratios dependency on the atmospheric parameters and its importance in the classification of irradiated exoplanets,} we aim to quantitatively investigate a 5D model parameter space to \revision{translate the photospheric chemistry into the spectra of irradiated planets.} We explore how these spectra change with the variation of planetary effective temperature, surface gravity, metallicity, carbon-to-oxygen ratio and spectral type of the host star. \revision{In this paper, as the first step, we present the results of our self-consistent cloud-free simulations, and in the forthcoming papers we address the effects of non-equilibrium chemistry and cloud opacities on these results to shape a consistent observationally driven theoretical framework on the classification of gaseous planets.}

\revision{In what follows,} we describe the most up-to-date version of our model (\emph{petitCODE}) and the parameter space that we have investigated in \hyperref[sec:method]{Section}\;\ref{sec:method}. In \hyperref[sec:results]{Section}\;\ref{sec:results}, we present the results on (C/O)\textsubscript{tr} ratios. In \hyperref[sec:discussion] {Section}\;\ref{sec:discussion}, we discuss how our chosen parameters influence the transition C/O ratios and propose a classification scheme for irradiated planetary spectra \revision{between 400 and 2600\;K} with four classes, and how they fit to Spitzer color-diagrams. We summarize and conclude our results and findings in \hyperref[sec:conclusion]{Section}\;\ref{sec:conclusion}.

\section{METHODS} \label{sec:method}
In order to investigate the influence of mentioned parameters on the atmospheric properties, we have synthesized a population of 28,224 self-consistent planetary atmospheres by using \emph{petitCODE} \citep{molliere_model_2015,molliere_observing_2017}. This code and the grid are described in the following subsections, and the grid is publicly available\footnote{\url{www.mpia.de/homes/karan}}.

\subsection{petitCODE} \label{subsec:petitCODE}


\emph{petitCODE} is a 1D model that calculates planetary atmospheric temperature profiles (TP structures), chemical abundances, and emission and transmission spectra (\revision{which includes} scattering), assuming radiative-convective and thermo-chemical equilibrium. It was introduced in \citet{molliere_model_2015}, and is described in its current form in \citet{molliere_observing_2017} (general capabilities) and \citep{molliere_detecting_2018} (opacity updates).

The basic physical inputs are the stellar effective temperature amd it radius, planetary effective temperature or distance to the star, planetary internal temperature, planetary radius and mass (or surface gravity), atomic abundances, and the irradiation treatment (it is possible to calculate the dayside average, planetary average, or to provide incidence angle for the irradiation).

There are two options to treat clouds: one is by following the prescription by \citet{ackerman_precipitating_2001} and introducing the settling factor ($f_{sed}$), the width of the log--normal particle distribution 
($\sigma_g$) and the atmospheric mixing $K_{zz}$; the other is by providing the cloud particle size and setting the maximum cloud mass fraction \citep[see][]{molliere_observing_2017}.

Depending on the case of interest, some of the inputs may not be required. For instance, the current \revision{paper presents our results on the} irradiated gaseous planets without clouds to provide a cloud-free framework, which can also be used to explore the atmosphere of cloud-free planets or as a diagnostic tool to identify cloudy/partially-cloudy atmospheres. \revision{As will be discussed in our following paper, our non-equilibrium chemistry study will be built upon this framework as well.} 

The chemical inputs of the code are the lists of atomic species (H, He, C, N, O, Na, Mg, Al, Si, P, S, Cl, K, Ca, Ti, V, Fe, and Ni) with their mass fractions and reaction products (H, \ce{H2}, He, O, C, N, Mg, Si, Fe, S, Al, Ca, Na, Ni, P, K, Ti, CO, OH, SH, \ce{N2}, \ce{O2}, SiO, TiO, SiS, \ce{H2O}, \ce{C2}, CH, CN, CS, SiC, NH, SiH, NO, SN, SiN, SO, \ce{S2}, \ce{C2H}, HCN, \ce{C2H2}, \ce{CH4}, AlH, AlOH, \ce{Al2O}, CaOH, MgH, MgOH, \ce{PH3}, \ce{CO2}, \ce{TiO2}, \ce{Si2C}, \ce{SiO2}, FeO, \ce{NH2}, \ce{NH3}, \ce{CH2}, \ce{CH3}, \ce{H2S}, VO, \ce{VO2}, NaCl, KCl, \ce{e-}, \ce{H+}, \ce{H-}, \ce{Na+}, \ce{K+}, \ce{PH2}, \ce{P2}, PS, PO, \ce{P4O6}, PH, V, VO(c), VO(L), \ce{MgSiO3(c)}, \ce{Mg2SiO4(c)}, SiC(c), Fe(c), \ce{Al2O3(c)}, \ce{Na2S(c)}, KCl(c), Fe(L), \ce{Mg2SiO4(L)}, SiC(L), \ce{MgSiO3(L)}, \ce{H2O(L)}, \ce{H2O(c)}, TiO(c), TiO(L), \ce{MgAl2O4(c)}, FeO(c), \ce{Fe2O3(c)}, \ce{Fe2SiO4(c)}, \ce{TiO2(c)}, \ce{TiO2(L)}, \ce{H3PO4(c)} and \ce{H3PO4(L)}) to be considered in the equilibrium chemistry network. The lists of gas opacity species (\ce{CH4}, \ce{H2O}, \ce{CO2}, HCN, CO, \ce{H2}, \ce{H2S}, \ce{NH3}, OH, \ce{C2H2}, \ce{PH3}, Na, K, TiO and VO) and cloud opacity species (\ce{Al2O3}, \ce{MgAl2O4}, \ce{Mg2SiO4}, \ce{MgSiO3}, \ce{MgFeSiO4}, Fe, KCl and \ce{Na2S}) should be also provided for the calculations. In addition, whether or not to include \ce{H2}-\ce{H2} collision induced absorption (CIA) and \ce{H2}-He CIA in the model must be specified.

\revision{The abundances in petitCODE follow from true chemical equilibrium, i.e. no “rain-out” of condensates is assumed \citep{burrows_chemical_1999,lodders_atmospheric_2002}. However, alkalis are not allowed to condense into feldspars, as Si atoms tend to be sequestered in rained-out silicates; see \citet{line_uniform_2017} for a discussion. In this way, the choice of allowed condensates effectively mimics the process of rain-out for the alkalis. This treatment of rain-out in petitCODE was found to be sufficient, as there is only very small differences found between the spectra and P-T structure solutions of petitCODE and Exo-REM \citep{baudino_toward_2017}, the latter of which includes rain-out.}

The code begins with an initial guess of the TP structure, that can be either user-provided or calculated from the \citet{guillot_radiative_2010} analytical solution.  The code then uses a self-written Gibbs-minimizer \citep[see][]{molliere_observing_2017}, resulting in chemical equilibrium abundances, as well as the adiabatic temperature gradient of the gas mixture. This chemical composition is then used to compute the opacities at each pressure level. Finally, the code computes the temperature profile assuming radiative-convective
equilibrium (both emission/absorption and scattering are taken into account, see paragraph below) and considers the new TP profile to iterate the procedure until convergence is reached. Finally, the code outputs emission and transmission spectra of the converged model at a resolution of $\lambda / \Delta \lambda = 1000$.

The temperature iteration method used in \emph{petitCODE} is a variable Eddington factor method. For this, the radiation fields of planet and star are both solved over the full wavelength domain (110~nm to 250~$\mu$m), for rays along 40 different angles (20 up and 20 down) with respect to the atmospheric normal. The angles $\vartheta$ are chosen for carrying out a 20-point Gaussian quadrature over $\mu = {\rm cos}(\vartheta)$. For the radiative transfer solution the Feautrier method is used. Scattering can be naturally included in the Feautrier method, and the scattering source function in \emph{petitCODE} is converged using both ALI \citep{olson_rapidly_1986} and Ng \citep{ng_hypernetted_1974} acceleration. In order to speed up calculations, scattering is assumed to be isotropic, but the scattering cross-sections are reduced by $(1-g)$, where $g$ is the scattering anisotropy factor. This ensures a correct scattering treatment in the diffusive limit \citep[see, e.g.,][]{wang_biomedical_2012}.

As reported in \citep{molliere_observing_2017}, the scattering implementation was tested by comparing the atmospheric bond albedo as a function of the incidence angle of the stellar light to the values predicted by Chandrasekhar's H functions \citep{chandrasekhar_radiative_1950}. For this test the appropriate simplifying assumptions were made (like vertically constant opacities). Excellent agreement was found. Because both the planetary and stellar radiation field are solved within the same, full wavelength regime, and along 40 rays, the radiative transfer is superior when compared to the often-used two-stream method. No assumptions have to be made for the direction that radiation is propagating into, or the wavelength range that the stellar or planetary radiation field typically populate, as long as both are within 110~nm to 250~$\mu$m. The only sense in which \emph{petitCODE} calculations may be considered as ``two-stream'' is the fact that they treat planet and stellar radiation independently. This in no way restricts the generality of the \emph{petitCODE} solutions, however, as the radiative transfer equation is linear in nature.

\emph{petitCODE} as described above, has been recently successfully benchmarked against the state-of-the-art \emph{ATMO} \citep{tremblin_fingering_2015} and \emph{Exo-REM} \citep{baudino_interpreting_2015} codes, see \citet{baudino_toward_2017}. Recent applications of the code include \citet{molliere_model_2015,mancini_optical_2016,molliere_observing_2017,southworth_detection_2017,baudino_toward_2017,samland_spectral_2017,tregloan-reed_possible_2017,muller_orbital_2018}, and \citet{molliere_detecting_2018}.

\subsection{Grid properties} \label{subsec:grid}
For modeling irradiated exoplanets, the main parameters of interest are typically the effective temperature (T\textsubscript{eff}), surface gravity (log(g)), metallicity ([Fe/H]), carbon-to-oxygen-ratio (C/O) and stellar type; for a recent review see \citet{fortney_modeling_2018}. In addition to these parameters, some other factors might be of significance as well, such as interior temperature, atmospheric thickness, eddy and molecular diffusion,
photochemistry and the presence of clouds. The effect of non-equilibrium chemistry and clouds will be presented in two follow-up papers where we will also introduce our \emph{Chemical Kinetic Model (ChemKM)} and \revision{our extensive self-consistent cloudy grid of models}.

The chosen parameters and their ranges are discussed in the following sub-sections.

\subsubsection{Effective temperature (T\textsubscript{eff})}\label{sec:Teff} 

Unless a planet is highly inflated, young, or far from its host star, the flux contribution from the planetary interior has a minimal effect on the atmospheric TP profile (e.g. \citealt{molliere_model_2015, fortney_modeling_2018}). This can be understood by a simple relation: T\textsubscript{eff}\textsuperscript{4}$=$T\textsubscript{eq}\textsuperscript{4}$+$T\textsubscript{int}\textsuperscript{4}, where T\textsubscript{eff}, T\textsubscript{eq} and T\textsubscript{int} are the effective, equilibrium and interior temperatures of the planet, respectively. In the limit of T\textsubscript{eff}$\gg$T\textsubscript{int},
the relation becomes T\textsubscript{eff}\textsuperscript{4}$\approx$T\textsubscript{eq}\textsuperscript{4}, and hence the flux contribution from the interior can be neglected.

We set the interior temperature at 200\;K to be consistent with the \citet{fortney_effect_2005} and \citet{molliere_model_2015} simulation setup so we can compare the effects of additional physics in the updated \emph{petitCODE} (such as extra reactants as well as including multiple scattering). Therefore the lowest T\textsubscript{eff} in our grid was selected to be 400\;K to keep the effect of interior temperature on the energy budget at a minimum.

To take a computationally pragmatic approach, we only studied planetary structures and spectra under the assumption of isotropic incident flux (i.e. planetary average). However, very hot planets are expected to display inefficient redistribution of the insolation energy to the night side due to the domination of radiative cooling over advection
\citep{perez-becker_atmospheric_2013,komacek_atmospheric_2016,keating_universal_2018}. For instance, Kepler-13Ab with T\textsubscript{eff}$\sim$2750\;K \citep{shporer_atmospheric_2014} and WASP-18b with T\textsubscript{eff}$\sim$3100\;K \citep{nymeyer_spitzer_2011} are shown to have low energy redistribution efficiencies, resulting in their large day-night temperature contrasts.  Therefore, we set the upper limit of T\textsubscript{eff} at 2600\;K, and investigate T\textsubscript{eff} from 400\;K to 2600\;K with an increment of 200\;K. \revision{This choice of temperature range also keep our parameter space away from the extremely irradiated hot Jupiters where the atmospheric chemistry is thought to be fundamentally different \citep{lothringer_extremely_2018}.}

\subsubsection{Surface gravity (log(g))}
\begin{figure}[t]
\includegraphics[width=\columnwidth]{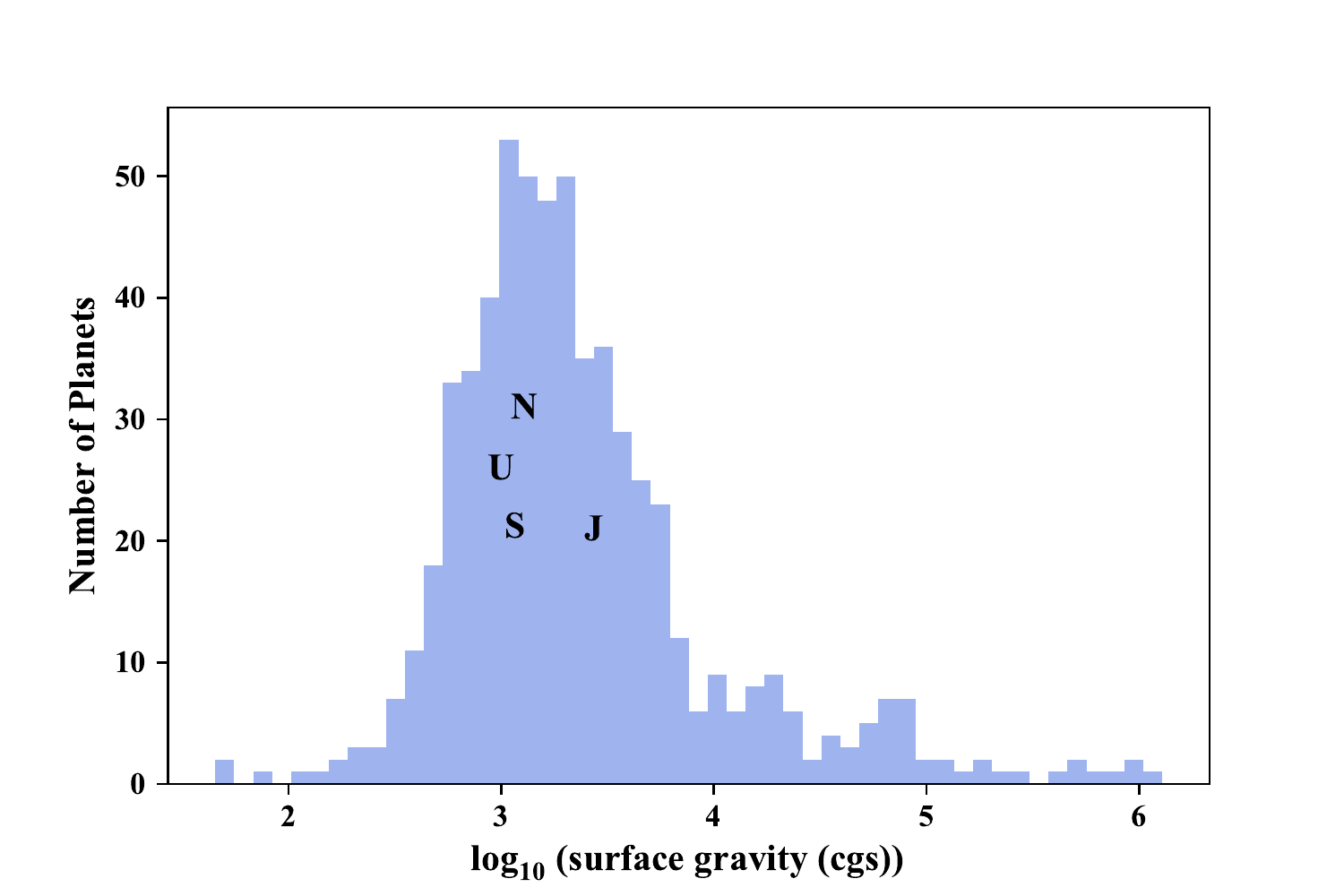}
\caption{Distribution of log(g) for known exoplanets. Values are estimated based on the retrieved radius and mass values from the NASA exoplanet archive. Log(g)s of the solar system’s gas-ice giants \revision{(J: Jupiter, S: Saturn, U: Uranus, and N: Neptune)} are also shown for comparison.\label{fig:logg}}
\end{figure}

In the solar system, radii of gas and ice giants are measured from the center up to an altitude where the pressure is 1 bar (see e.g. \citealt{simoes_schumann_2012,kerley_structures_2013,robinson_common_2014}). For exoplanets we determine the radius by photometry and estimating where their atmospheres become optically thick. This radius is called the \emph{photospheric radius}. Applying the photometric approach on the Solar System\textsc{\char13}s gas and ice giants does not provide us with the same values, but it is still a valid approach to estimate their radii. Discrepancies of the two methods however remain within a few percent.

Nevertheless, in most cases the mass or the radius of an exoplanet are not well constrained and one can use surface gravity as a combined quantity to explore the effect of these two quantities by one combined parameter (see e.g.
\citealt{fortney_modeling_2018}). In addition, the temperature structure calculations depend only on the surface gravity and not the planetary radius and mass as two separated parameters. Therefore, the selection of surface
gravity over radius and mass of the planet remains a plausible choice.

\hyperref[fig:logg]{Figure}\;\ref{fig:logg} illustrates the distribution of log(g) based on radius and mass values retrieved from the \emph{NASA exoplanet archive}
\footnote{\href{https://exoplanetarchive.ipac.caltech.edu/index.html}{exoplanetarchive.ipac.caltech.edu}}. Log(g) ranges from 1.5 to 6.1 with only a few objects at the extreme values. Note that high log(g) values are mostly associated with the objects having masses larger than 13M\textsubscript{Jupiter} and hence, by definition (see e.g. \citealt{homeier_spectral_2005}), are Brown dwarfs. We thus explored this parameter from 2.0 to 5.0 with increment of
0.5.

\subsubsection{Metallicity {\normalfont([Fe/H])}}
The metallicities of solar system gaseous planets range from \revision{around 3 times} to 100 times of solar metallicity. There is a trend of higher metallicity for lower-mass objects. Observations suggest that this trend holds true for exoplanets as well \citep{miller_heavy-element_2011,thorngren_massmetallicity_2016,wakeford_complete_2017,sing_observational_2018}, but it should be kept in mind that this conclusion is only based on a few estimations with large uncertainties. Furthermore the metallicities of these different exoplanets have been estimated using different definitions and techniques, and thus it is difficult to make a fair comparison between them \citep{heng_what_2018}.

Nevertheless, we chose to explore a wide range of metallicities from sub-solar, [Fe/H]=-1.0, to super-solar [Fe/H]=2.0 with increment of 0.5. [Fe/H] denotes the metallicity in log-scale where [Fe/H]=-1.0 represents an atmosphere with 10 times lower \emph{metal} abundances than in the Sun; here \emph{metal} refers to all elements except H and He.

\subsubsection{Carbon-to-oxygen-ratio {\normalfont(C/O)}}\label{sec:C_O} 

As briefly discussed in the introduction, varying C/O alters the TP structure as well as the abundance distribution of species in the atmosphere. The highest sensitivity of TP and chemical abundances to C/O variations is expected to occur around C/O$\sim$1 where the natural boundary between methane- and water-dominated atmospheres is predicted and reported. For this reason we selected irregular parameter steps spanning from 0.25 to 1.25 with smaller steps around unity: C/O=[0.25, 0.5, 0.7, 0.75, 0.80, 0.85, 0.90, 0.95, 1.0, 1.05, 1.10, 1.25]. Unlike the definition of metallicity, C/O represents the number ratio of carbon to oxygen elemental abundances and is not scaled to the solar value of $\sim$0.55.

\revision{In principle, there are three ways to alter C/O ratio: by changing the oxygen abundance but keeping the carbon abundance fixed, by changing the carbon abundance but keeping the oxygen abundance fixed, and changing both but keeping the total oxygen and carbon abundance constant. The compositional outcome of these scenarios can be quite different. \citet{lodders_exoplanet_2010} discussed the first two scenarios and reported the different compositional outcome of these two cases. Changing the oxygen abundance (to alter C/O ratio) represents the accretion of gas or planetesimals with different water contents onto a forming planet. Similar to \citet{madhusudhan_c/o_2012,molliere_model_2015} and \citet{woitke_equilibrium_2018}, we also follow this school of thought.}

\subsubsection{Stellar type}

Irradiated atmospheres are susceptible to their parent star\textsc{\char13}s spectral type. As the temperature of the host increases its spectral peak\textsc{\char13}s wavelength decreases toward the blue region of the spectrum, affecting the optically active parts of the planetary atmospheres. The effect of stellar spectral type on irradiated
atmospheres has been investigated by a number of authors (see e.g. \citealt{miguel_effect_2014,molliere_model_2015,fortney_modeling_2018}). In this work we chose the same values for this parameter as in \citet{molliere_model_2015}, i.e. M5, K5, G5 and F5, to cover a wide range of stellar types and make the models directly comparable with their grid of models.

\subsubsection{Reactants and Opacity sources} 

We kept all \emph{petitCODE}’s atomic species and reaction products (including TiO/VO) in our models except one reactant, \ce{MgAl2O4(c)}, due to the poor convergence of some of the models. We discuss this common problem in the forward models and our solution to it in \hyperref[sec:problem_petit]{Appendix}\;\ref{sec:problem_petit}. We considered these gas opacity species: \ce{H2O}, CO, \ce{CO2}, OH (HITEMP, see \citealt{rothman_hitemp_2010}), \ce{CH4}, HCN (ExoMol, see \citealt{tennyson_exomol:_2012}), as well as \ce{H2}, \ce{H2S}, \ce{C2H2}, \ce{NH3}, \ce{PH3} (HITRAN, see \citealt{rothman_hitran2012_2013}), Na, K (VALD3, see \citealt{piskunov_vald:_1995}) and \ce{H2}-\ce{H2} and \ce{H2}-\ce{He} CIA \citealt{borysow_collision-induced_1989,borysow_collision-induced_1989-1,richard_new_2012}), but no cloud opacity. We shall present and discuss the effects of TiO/VO and cloud opacities on planetary atmospheres in a follow-up paper.

\section{RESULTS} \label{sec:results}
Given our grid setup and parameters of choice, we calculated 28,224 forward self-consistent models of planetary atmospheres and their transmission and emission spectra. For calculating the transmission spectra we set the reference pressure 1 R\textsubscript{jup} at 10 bar, following \citet{fortney_transmission_2010} prescription. In order to quantitatively discriminate spectral features and how they vary from one spectrum to another, we introduced a technique to decompose a spectrum to its individual opacity sources. This technique is discussed in the following section.

\subsection{Spectral Decomposition Technique} \label{sec:decomposition}

\begin{figure*}
\includegraphics[width=\textwidth]{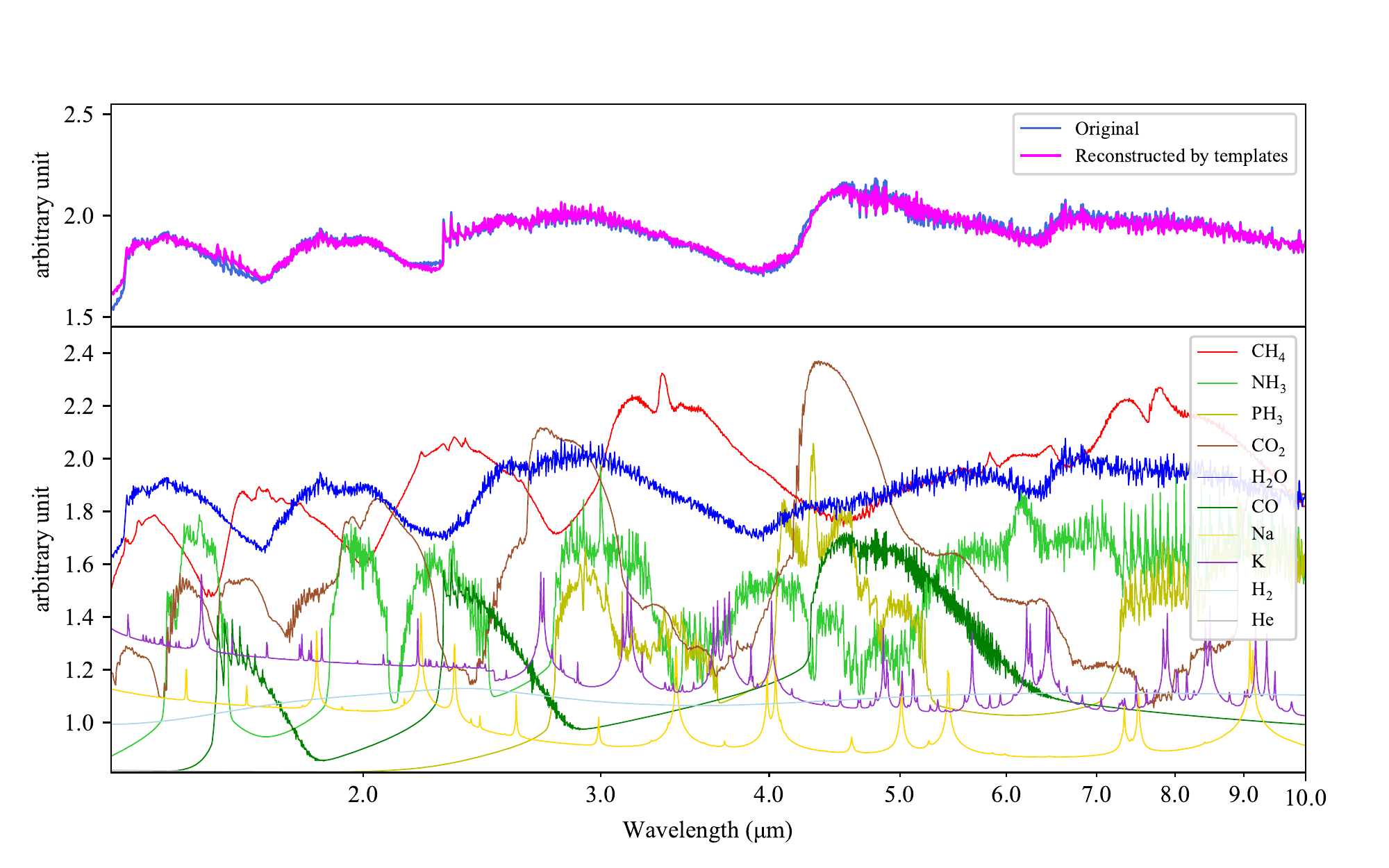}
\caption{Reconstruction of a transmission spectrum by the \emph{Spectral Decomposition Technique}. Top) Transmission spectrum of a planet with an effective temperature of 1600\;K, \revision{log(g)=3.0, [Fe/H]=1.0, C/O=0.85 orbiting a central G5 star} (blue) and its reconstruction by the Spectral Decomposition Technique (magenta). Bottom) The templates of major opacity sources in the reconstructed model. \label{fig:fig4}}
\end{figure*}

Thermal emission at any given wavelength comes from a range of pressures, but the contribution of emission flux from each pressure level in the final emergent emission spectrum is not equal. A common practice is to define a \emph{contribution function} and evaluate how sensitive the emission spectrum is to different pressure levels (see e.g. \citealt{selsis_search_2002,swain_water_2009,molliere_modeling_2017,cowan_mapping_2017,dobbs-dixon_wavelength_2017,fortney_modeling_2018}). Similarly, the contribution function can be calculated for the transmission spectra.

While this method combines information of all atmospheric constituents to provide the spectral contribution at each pressure, another approach could be taken to define a \emph{contribution coefficient} for each species to approximate its contribution in the spectrum, integrated over all pressures. This would allow us to study the relative importance of individual species in a given spectrum and to investigate the dominant net chemical reactions at the photospheric levels of the planet that cause those spectral signatures. We call this method the \emph{Spectral Decomposition Technique} and develop it for the decomposition of transmission spectra.

This technique was motivated by the fact that opacities contribute logarithmically to the transmission spectra, and major atmospheric opacity sources (such as \ce{H2O}, \ce{CO2}, \ce{NH3}, \ce{CH4}, \ce{HCN}, \ce{CO}, \ce{C2H2}) have distinct signatures in the range of optical to IR wavelengths:
\begin{equation} 
    z(\lambda)=H_p\:\log\Big[\sum_{i} \kappa_i (\lambda)\Big] +
cst,  
\label{eq:eq_trans} 
\end{equation}
\citep{fortney_effect_2005,lecavelier_des_etangs_rayleigh_2008}, where $z(\lambda)$ is the photometric radius at the wavelength $\lambda$, $H_p$ is the atmospheric scale height and $\kappa_i$ is the opacity of the $i^{th}$ species above a reference pressure (or a reference radius, interchangeably). This technique has been employed before, but only graphically. For example, in a study of hot-Jupiters spectra by \citet{rocchetto_exploring_2016}, they provided several synthetic transmission spectra along with the contributions of the major opacity sources to illustrate how much they contribute to the spectrum qualitatively. For additional examples see \citet{tinetti_exploring_2010,shabram_transmission_2011,encrenaz_transit_2015} and \citet{kreidberg_exoplanet_2017}.

The first step to decompose a spectrum to its individual opacity sources is to produce a template of every species, $\Gamma_i (\lambda)$. For a transmission spectrum, this can be achieved by assuming the template spectra to contain only a given species each; e.g. the water template has only \ce{H2O} in the atmosphere and the methane template has only \ce{CH4} and so on. The TP profile in the templates can be adopted directly from the self-consistently calculated TP structure of each model. However, if the decomposition is intended for an extensive number of \revision{transmission spectra}, a reasonable approximation would be to employ an isothermal TP and calculate the templates only once. Here we followed the latter and set the temperature at 1600\;K for the calculation of templates. It is then possible to estimate the contribution coefficient of each opacity source, $c_i$, using \hyperref[eq:eq8]{Equation}\;\ref{eq:eq8}. 
\begin{equation}
    \mathcal{S}(\lambda)=\Big(\sum_{i} c_i 
\Gamma_i^p(\lambda)\Big)^\frac{1}{p},
\label{eq:eq8}
\end{equation}
where $\Gamma_i (\lambda)$ is the spectral template of the $i^{th}$ species, p is an arbitrary exponent that can be adjusted to achieve the best result over a wide range of parameter space (here we chose it to be 10; higher values make stronger spectral features more pronounced), and $\mathcal{S}$ is the total spectrum.

After creating a spectral template for each opacity source, as shown for example in \hyperref[fig:fig4]{Figure}\;\ref{fig:fig4}, bottom panel, we can raise the templates to the p\textsuperscript{th} power, multiply them by some coefficients, add them up and then take the p\textsuperscript{th} root of the summed spectrum to calculate the total spectrum, $\mathcal{S}$. We explored different combinations to find the best linear combination of the templates that could represent the spectrum, \hyperref[fig:fig4]{Figure}\;\ref{fig:fig4}, top panel. The coefficients of this best linear combination are the \emph{contribution coefficient} of species.

\revision{To perform the decomposition, a wavelength range should be chosen. Using wider wavelength ranges generally results in a more accurate estimation of contribution coefficients; However, depending on the species of interest not all wavelengths have the same information content. In the current study, the aim is to estimate the transition C/O ratios by the use of \ce{H2O} and \ce{CH4} contribution coefficients. Therefore a choice of 1.3-10\;$\mu m$ sufficiently provides the spectral information content needed for the spectral decomposition to achieve the same results as a choice of 0.4-20\;$\mu m$, the latter of which is the wavelength span of our synthetic spectra.}

\revision{While not the focus of our current study, one could similarly perform the spectral decomposition on a cloudy transmission spectrum. Since clouds and hazes may obscure or mute spectral features, it is therefore important to introduce a template for the cloud/haze species to estimate the contribution coefficients. This will be discussed in a forthcoming paper describing our self-consistent cloudy grid.}

The illustrated example in \hyperref[fig:fig4]{Figure}\;\ref{fig:fig4} presents a case with an effective temperature of 1600\;K, \revision{log(g)=3.0, [Fe/H]=1.0, C/O=0.85 and G5} to be the host star’s spectral type. The transmission spectrum (blue curve in the top panel) shows clear signatures of both \ce{H2O} and CO molecules between 1-4\;$\mu$m. Small excess absorption at longer wavelengths, and particularly at $\sim$4.2\;$\mu$m, might hint the presence of \ce{CO2}, but \ce{CH4} has almost no contribution in the spectrum. By using the spectral decomposition technique the contribution coefficients of \ce{H2O}, \ce{CH4}, CO and \ce{CO2} were found to be \revision{0.88, 0.03, 1.25 and 0.08}, respectively, consistent with our visual interpretation of the spectrum.

The ratio of contribution coefficients provides a quantitative estimation of spectral contrast for different species as a measure of species’ relative detectability. For instance as $c_{\ce{CH4}}/c_{\ce{H2O}}$  increases, methane features become more pronounced in the spectrum with respect to the water features and hence the probability of \ce{CH4} detection increases. This ratio can then be used to determine the dominance of observable atmosphere by water or methane. In the next section we show how to apply this method on the models with different C/O ratios in order to estimate the \emph{transition} C/O ratios.

\subsection{Estimation of transition C/O ratios} \label{subsec:C/Os}

\begin{figure*}
\includegraphics[width=\textwidth]{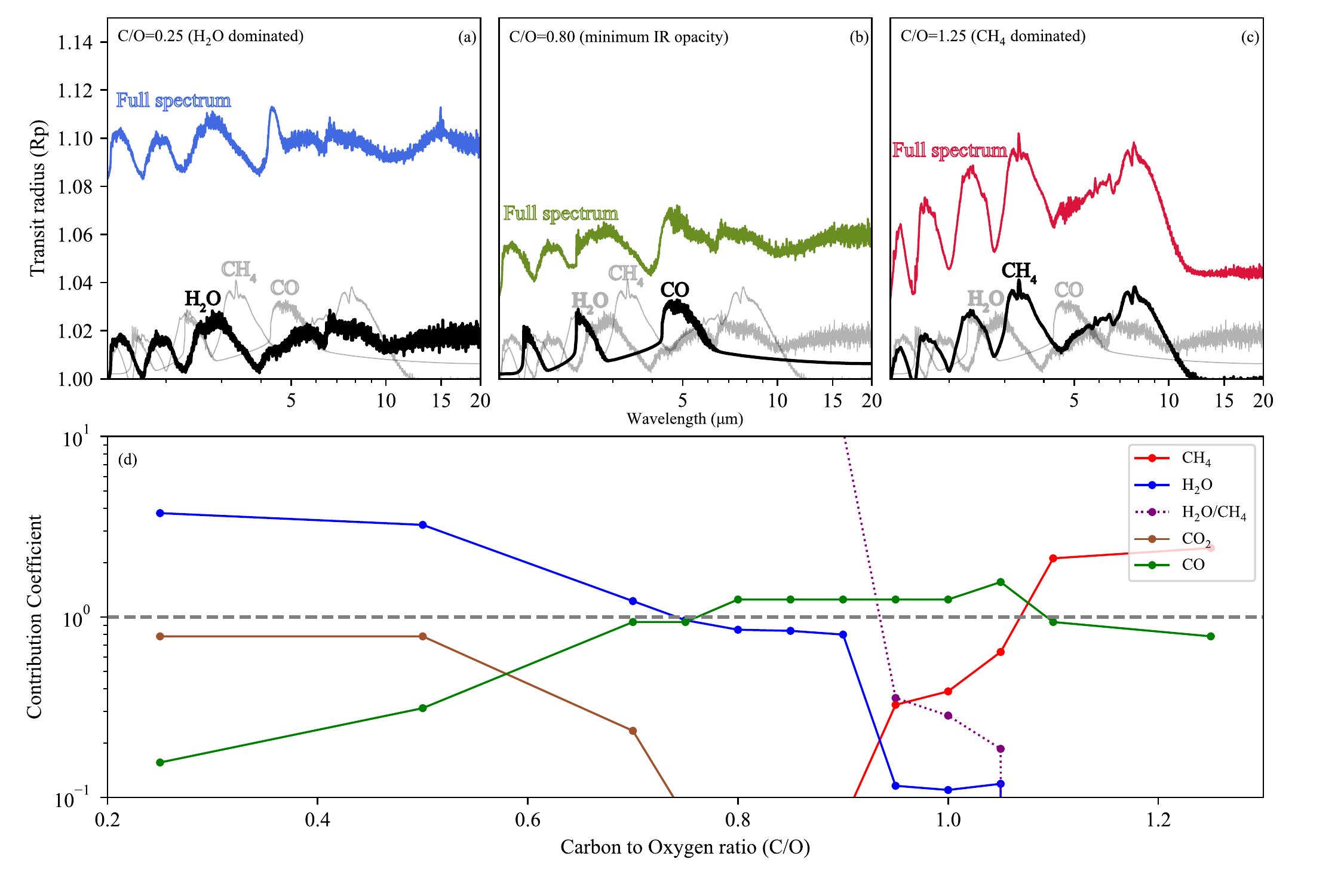} \caption{Top panels) Transition of water- to methane-dominated transmission spectrum for a planet with \revision{T\textsubscript{eff}$=$1600\;K, log(g)=3.00, [Fe/H]=1.00 orbiting an G5} star. \revision{(a) At low C/O ratio, \ce{H2O} dominates the spectrum (blue). (b) Higher C/O ratio decreases \ce{H2O} abundances and hence the transit radius decreases (green). (c) By increasing C/O ratio \ce{CH4} abundance increases and hence contributes more in the spectrum and the transit radius increases (red). Black and gray lines represent \ce{H2O}, \ce{CO}, and \ce{CH4} templates with an arbitrary offset to graphically illustrate which of them have a higher contribution into the spectrum (black).} Bottom) Variation of contribution coefficients of \ce{H2O}, \ce{CH4}, \ce{CO2} and \ce{CO} at different C/O ratios. As the ratio of \ce{H2O} to \ce{CH4} contribution coefficient (dotted purple line) passes the unity (gray horizontal dashed line), the spectrum transitions from water- to methane-dominated one; \revision{resulting in a quantitative estimation of (C/O)\textsubscript{tr}.} \label{fig:fig5}}
\end{figure*}

\begin{figure*}
\includegraphics[width=\textwidth]{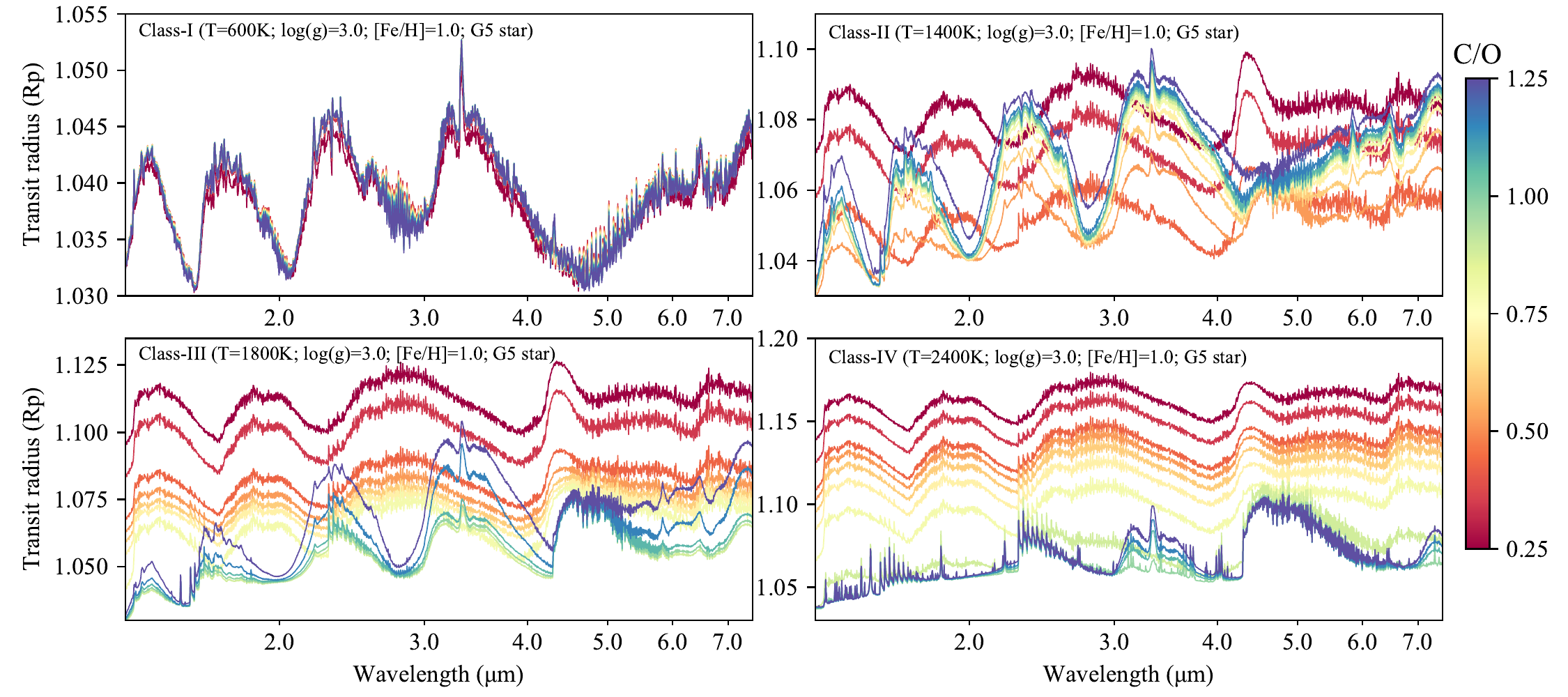}
\caption{Examples of spectral variation in four planetary \emph{Classes} as C/O increases from 0.25 (blue) to 1.25 (red), with the steps described in \hyperref[sec:C_O]{Section}\;\ref{sec:C_O}. In \emph{Class-I} \revision{(upper left)}, higher C/O ratios result in slightly more \ce{CH4} abundant atmospheres but water features are also remain noticeable. More \ce{CH4} also cause the planets to appear larger and a transition from \ce{H2O}- to \ce{CH4}- dominated spectrum does not coincide with a minimum IR opacity condition as it is the case for the \emph{Class-II} \revision{(upper right)}. In \emph{Class-II} and \emph{Class-III} \revision{(lower left)} the transition occurs at the minimum IR opacity, where the planet appears to be smaller relative to other similar cases but with different C/O ratio. In \emph{Class-IV} \revision{(lower right)}, a higher C/O results in a smaller radius for planets since larger C/O ratio leads to a stronger removal of water from the atmosphere, but the condition is in favor of \ce{CH4} destruction and hence there is no significant opacity sources in IR to make the
planet larger. \label{fig:paper1_spectrum_exmaples_four_classes}} \end{figure*}

\begin{figure*}
\plotone{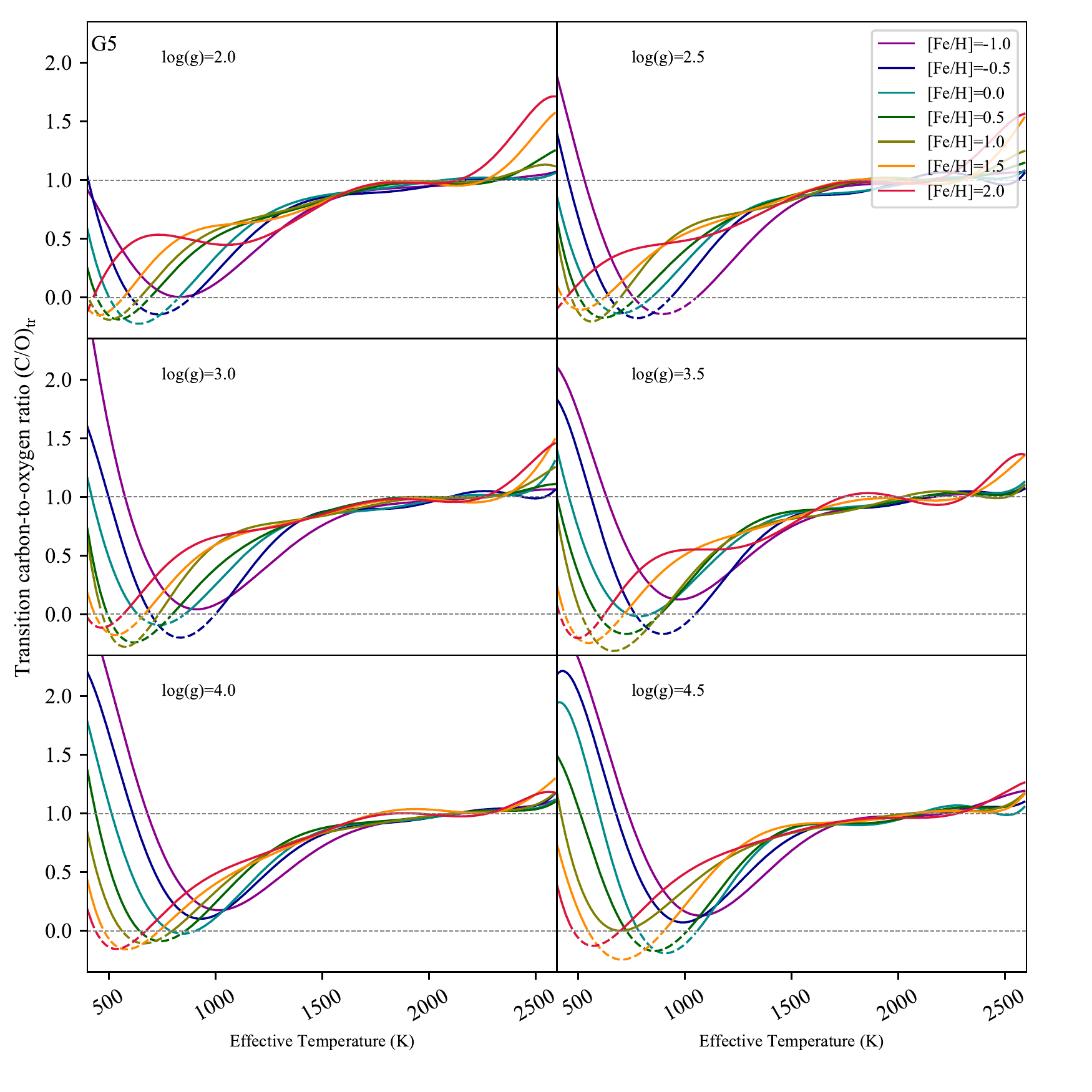}
\caption{Transition carbon-to-oxygen ratios, (C/O)\textsubscript{tr}, for planets around a G5 star. (C/O)\textsubscript{tr} marks the condition where the atmospheric spectrum transits from water- (regions under the
transition lines) to methane-dominated features (regions above the transition lines). Sub-panels show the values at different surface gravities, log(g). (C/O)\textsubscript{tr} values below 0.25 and above 1.25 are outside of our parameter range and are linearly extrapolated from the neighboring points.
$(C/O)_{tr}\leq 0$ means for that particular log(g) and [Fe/H] the spectrum is predicted to be always methane-dominated regardless of C/O ratio, assuming equilibrium chemistry (colored dashed curves below C/O=0). \label{fig:fig6}}
\end{figure*}

Following our previous example of a planet with an effective temperature of 1600\;K, \revision{log(g)=3.0, [Fe/H]=1.0 and a central G5} star, we explore a variety of C/O ratios and estimate the contribution coefficients of major opacity species. At C/O=0.25 the \ce{H2O}, \ce{CH4}, CO and \ce{CO2} contribution coefficients are equal to 3.8, 0.0, 0.14 and 0.79 respectively: a clear indication of a water-dominated spectrum and no trace of \ce{CH4}, see \hyperref[fig:fig5]{Figure}\;\ref{fig:fig5}a. At C/O=0.5, these coefficients change to 3.2, 0.0, 0.31 and 0.79, suggesting more CO and less \ce{H2O} spectral contributions. The trend continues at C/O=0.7 with 1.23, 0.0, 0.93 and 0.23 values for the coefficients. In all of these models, \ce{CO2} closely follows the water features’ diminishing trend, i.e. \ce{CO2}\textsc{\char13}s contribution coefficient is strongly correlated with \ce{H2O}\textsc{\char13}s contribution coefficient; brown and blue curves in \hyperref[fig:fig5]{Figure}\;\ref{fig:fig5}d respectively, implying they are both part of the same net chemical reaction.

\ce{CH4} begins to contribute at C/O=0.90 by $c_{\ce{CH4}}=0.02$, see \hyperref[fig:fig5]{Figure}\;\ref{fig:fig5}b, and at C/O=0.95 its contribution surpasses water\textsc{\char13}s; leading to a methane dominated spectrum. A linear interpolation suggests the transition occurs at (C/O)\textsubscript{tr}=0.96 where $c_{\ce{CH4}}/c_{\ce{H2O}}=1$; consistent with the value reported by \citet{molliere_model_2015} for the planets with T\textsubscript{eff} $\geq$ 1750\;K. At this transition C/O ratio, both water and methane opacities contribute very little to the spectrum and carbon monoxide has the highest contribution, see the green curve in \hyperref[fig:fig5]{Figure}\;\ref{fig:fig5}d. CO is not a significant IR opacity source compared to \ce{H2O} and \ce{CH4} and therefore diminished contributions of \ce{H2O} and \ce{CH4} result in a minimum atmospheric IR opacity such that an inversion is expected to form for hot planets with host stars of type K and earlier \citep{molliere_model_2015}. Equilibrium chemistry maintains methane\textsc{\char13}s spectral dominance at all higher C/O ratios for the case that we studied here, \hyperref[fig:fig5]{Figure}\;\ref{fig:fig5}c,d.

Decomposing the spectra for a similar case but with lower metallicity [Fe/H]=-1.0 results in a lower transition C/O ratio of 0.83 relative to the case of [Fe/H]=1.0. This can be understood by considering \hyperref[eq:eq5]{Equation}\;\ref{eq:eq5} for relatively hot planets, \citep{molliere_model_2015}:
\begin{equation}
    \frac{d\gamma}{d[Fe/H]}>0,
\label{eq:eq5}
\end{equation}
where $\gamma$$=$$\kappa$\textsubscript{vis}$/$$\kappa$\textsubscript{IR} and $\kappa$\textsubscript{vis} and $\kappa$\textsubscript{IR} are the mean opacities in the visual and IR wavelengths in the atmosphere, respectively. Therefore, the cooling efficiency of the atmosphere is expected to increase as [Fe/H] decreases. A colder environment, in turn, is in favor of more \ce{CH4} production and thus the transition occurs at lower C/O ratios in this case.

The spectral dominance of methane features over water features does not mean a complete lack of water features in the spectrum, but rather it is the relative strength of methane features in comparison to the water features. For instance, exploring somewhat colder planets (T\textsubscript{eff}$\lesssim$1000\;K) reveals that both water and methane features are present in the spectra, see e.g. T\textsubscript{eff}$=$600 case in \hyperref[fig:paper1_spectrum_exmaples_four_classes]{Figure}\;\ref{fig:paper1_spectrum_exmaples_four_classes}.

Calculating the transition C/O ratios for all 28,224 models reveals similar trends to the predicted trends by \citet{molliere_model_2015}. As an example, \hyperref[fig:fig6]{Figure}\;\ref{fig:fig6} shows calculated transition C/O ratios for planets around a G5 star. Although, the general trend remains similar to the prediction by \citet{molliere_model_2015}, the details of the trends differ for different log(g) and [Fe/H] values. We extrapolated the transition C/O ratios when they occurred outside of our C/O parameter range, i.e. C/O$<$0.25 or C/O$>$1.25. Because of this, it is possible to also numerically find (C/O)\textsubscript{tr}$<$$0$. These negative
(C/O)\textsubscript{tr} values indicate the parameter space where the spectrum is expected to be always methane-dominated and has no other physical interpretation; colored dashed curves below C/O=0 in \hyperref[fig:fig6]{Figure}\;\ref{fig:fig6} show these regions. Negative ratios notwithstanding, we draw the extrapolated trends to aid the eye since the location of minimum (C/O)\textsubscript{tr} in this temperature range is key to separate the
first two atmospheric classes as will be discussed in the next section.

\section{Discussion} \label{sec:discussion}

\subsection{Four classes of atmospheric spectra} \label{subsec:classes}

Trends of (C/O)\textsubscript{tr} values at different temperatures suggest four
classes of distinct chemically driven planetary spectra \revision{in a cloud-free context}. Hence we propose a spectral classification scheme of irradiated planets based on these classes \revision{as a preparatory step to comprehend an observationally driven classification scheme with additional physics. Follow-up studies are needed to confirm and refine this classification framework.}

The first class contains cold planets with T\textsubscript{eff} lower than
$\sim$600-1100\;K. Their (C/O)\textsubscript{tr} ratios have a quasi-linear
relation with the effective temperature, i.e. for a given metallicity and
surface gravity, (C/O)\textsubscript{tr} linearly decreases as temperature
increases. This can be traced back to the dominant net reactions in this temperature range \citep{pirie_manufacture_1958,atreya_origin_1989}:
\begin{equation}
\label{eq:eq9}
\ce{CO + 3H2 -> H2O + CH4 }
\end{equation}
\begin{equation}
\label{eq:eq10}
\ce{CO + H2O -> CO2 + H2 }
\end{equation}
where oxygen and carbon atoms are mostly bond in water and methane molecules,
but \ce{CO2} can lock up a fraction of oxygen atoms, too. Since \hyperref[eq:eq9]{Reaction}\;\ref{eq:eq9} is strongly pressure sensitive, the chemical equilibrium abundances are thus highly temperature and pressure dependent. Consequently (C/O)\textsubscript{tr} ratios are expected to change significantly, depending on the metallicity and surface gravity of the planet.
This can be noticed in the diversity of (C/O)\textsubscript{tr} values at low temperatures. We stress again that both \ce{CH4} and \ce{H2O} features are expected to be present in the spectra of these planets since the overall temperature-pressure at the photospheric level of this class
favors production of both \ce{CH4} and \ce{H2O}, see \emph{Class-I} in
\hyperref[fig:paper1_spectrum_exmaples_four_classes]{Figure}\;\ref{fig:paper1_spectrum_exmaples_four_classes}. \revision{In reality, however, non-equilibrium chemistry and cloud formation are expected to obscure or mute some of the spectral features in the spectra of this class (see e.g. \citet{sing_continuum_2016}). Since photosphere of planets with higher metallicity and lower surface gravity extends to lower pressures, the spectra of this kind of class-I planets are expected to be quite vulnerable to the non-equilibrium chemistry and presence of clouds. We will examine this prediction in the forthcoming papers.}

The second class contains intermediate-temperature planets, i.e.
T\textsubscript{eff} higher than \emph{Class-I} but lower than $\sim$1800\;K.
For this class, (C/O)\textsubscript{tr} highly depends on the surface gravity
and metallicity, see \hyperref[fig:fig6]{Figures}\;\ref{fig:fig6} and
\ref{fig:fig8}. The main net reaction is similar to the dominant chemical
reaction in \emph{Class-I} but toward the other direction due to higher
temperatures. Therefore:
\begin{equation}
\label{eq:eq11}
\ce{H2O + CH4 -> CO + 3H2 }
\end{equation}

where the condition is in favor of CO production. Due to the presence of oxygen-containing condensates in this temperature range, the transition of water-to-methane-dominated-spectra depends on how much condensates are evaporated, which in turn depends on the metallicity and log(g) of the planet. \revision{Theoretical predictions (see e.g. \citet{ackerman_precipitating_2001,fortney_effect_2005,helling_cloud_2008,moses_disequilibrium_2011,heng_understanding_2013,venot_chemical_2015,wakeford_transmission_2015,drummond_effects_2016,kempton_observational_2017}) and observations (see e.g. \citet{madhusudhan_high_2011,knutson_3.6_2012,sing_continuum_2016}) suggest that non-equilibrium chemistry and clouds could be present even in hotter exoplanets, although less likely comparing with class-I \citep{wakeford_transmission_2015, stevenson_quantifying_2016, wakeford_high-temperature_2017}. These can also potentially alter the oxygen and carbon abundances in this class, and as a result, the dominant chemistry at the photospheric levels and therefore the spectra can change as well. We will briefly discuss the observational evidence in the next section.}

As the effective temperatures of the planets increase, condensates are completely evaporated. The net \hyperref[eq:eq11]{Reaction}\;\ref{eq:eq11}
still dominates the chemical equilibrium but the lack of silicates and other oxygen carrier condensates from the spectrally active regions of the atmosphere at these temperatures, T\textsubscript{eff}$>$1800\;K, forces the transition to be almost independent of log(g) and [Fe/H] \citep{molliere_model_2015}. Therefore, (C/O)\textsubscript{tr} remains at around a constant value, see \hyperref[fig:fig6]{Figure}\;\ref{fig:fig6} and \hyperref[fig:fig8]{Figure}\;\ref{fig:fig8}, and \emph{Class-III} of planets emerges. \revision{Although the presence of clouds is expected to be less probable for this class, due to the lack of condensates, the importance of dynamics and cooling mechanisms on the nightside of these planets can not be neglected. Therefore, clouds and out-of-chemical-equilibrium atmospheric constituents can be transported to the dawn terminator from the nightside and alter the transmission spectrum, but the dayside emission spectrum is likely to remain unaffected.}

At even higher temperatures, i.e. \emph{Class-IV} with
T\textsubscript{eff}$>$2200\;K, HCN dominates the atmosphere as the main carbon-bearing compound through three possible net reactions, also see \citet{molliere_model_2015}:
\begin{equation}
\label{eq:eq12}
\ce{CH4 + NH3 -> HCN + 3H2 }
\end{equation}
\begin{equation}
\label{eq:eq13}
\ce{2CH4 + N2 -> 2HCN + 3H2 }
\end{equation}
\begin{equation}
\label{eq:eq14}
\ce{NH3 + CO -> H2O + HCN }
\end{equation}
Bimolecular reaction rates of \hyperref[eq:eq12]{Reaction}\;\ref{eq:eq12} increase by one order of magnitude from 700 to 1400\;K, at around one millibar \citep{hasenberg_hcn_1987} and the condition at high temperatures progresses in favor of \ce{CH4} and \ce{NH3} destruction as well as HCN production. This results in the reappearance of the (C/O)\textsubscript{tr} dependency on log(g) and [Fe/H] and a mild increase in (C/O)\textsubscript{tr} at higher temperatures that will be discussed in the following section.

Altogether, four spectral classes can be defined based on their dominant chemical reactions and major IR spectral characteristics \revision{within the parameter space of this study}. \hyperref[fig:paper1_spectrum_exmaples_four_classes]{Figure}\;\ref{fig:paper1_spectrum_exmaples_four_classes}
shows some example spectra in each \emph{Class} where the dominant spectral features change as C/O ratios increase from 0.25 to 1.25 (blue to red colors in the Figure).

Three out of four classes, i.e. first, second and fourth classes, show dependency of the transition (C/O)\textsubscript{tr} ratios on log(g) and [Fe/H] which is discussed in the next section.

\subsection{Effect of {\normalfont log(g)} and {\normalfont[Fe/H]}} \label{subsec:classes_log_feh}

\begin{figure*} \includegraphics[width=\textwidth]{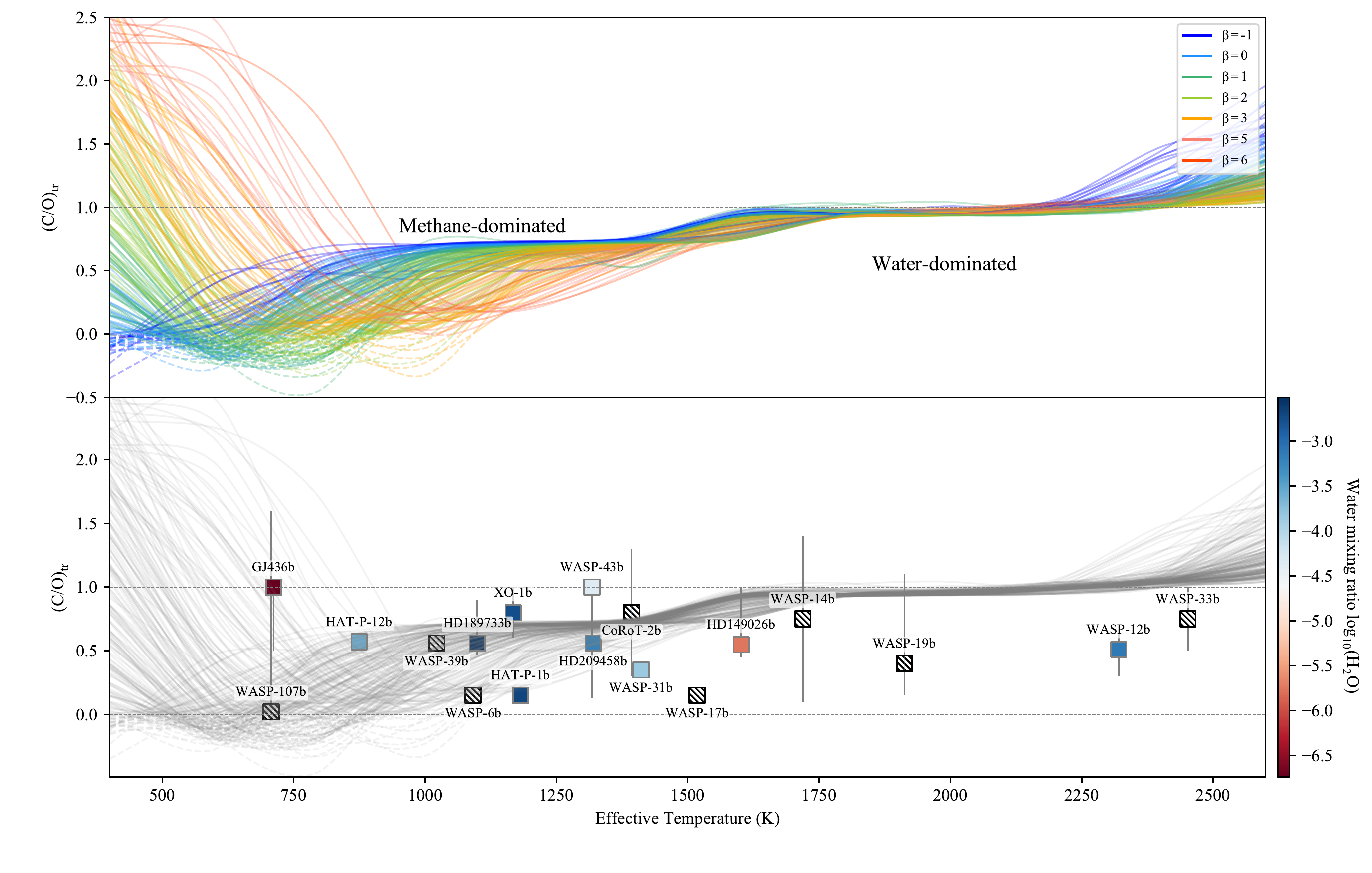} \caption{ Top) Mapping all water-to-methane transition curves reveals a region between 800 and 1500\;K with C/O$>$0.7 (\emph{the Methane Valley}) where methane spectral features are always the dominant features in the planetary spectra, given the assumptions used in this study. Exploring this region could enhance the probability of methane detection. The lack of such detection could alternatively provide a suitable road-map to study out of thermo-chemical, out of radiative-convective hydrostatic equilibrium, or the effect of clouds on the presence of water/methane abundances in planetary atmospheres. \revision{Bottom) Estimated C/O ratios for several planets tentatively suggest departure from cloud-free equilibrium chemistry conditions since the observed planets within or close to this region (i.e. WASP-43b, XO-1b, HD 189733b, and HAT-p-12b) all contain significant water abundances in their photosphere \citet{tsiaras_population_2018}. More precise C/O measurements are needed to observationally constrain the Methane Valley properties.}\label{fig:fig8}} \end{figure*}

At any given temperature of the first class, increasing the metallicity
decreases (C/O)\textsubscript{tr} ratio, see
\hyperref[fig:fig6]{Figures}\;\ref{fig:fig6} and \ref{fig:fig8}. This can be
understood by considering the net Reactions\;\ref{eq:eq9} and\;\ref{eq:eq10}.
By combining those two reactions we arrive at a new net reaction:
\begin{equation}
\label{eq:eq15}
\ce{CO2 + 4H2 <=> 2H2O + CH4 }
\end{equation}
The net \hyperref[eq:eq15]{Reaction}\;\ref{eq:eq15} establishes a one-to-one
relation between \ce{H2O} and \ce{CO2} where it favors \ce{CO2} production at
high pressure and high metallicity conditions. As metallicity increases, oxygen
can be locked up in \ce{CO2} more readily in comparison to CO 
(see e.g. \citealt{heng_carbon_2016}). This enhances the reduction of water abundance at the photosphere and results in a decreased (C/O)\textsubscript{tr} ratio.

Likewise, decreasing the surface gravity decreases the (C/O)\textsubscript{tr} ratio. A simple relation
between the optical depth ($\tau$) and pressure ($P$) of a planetary atmosphere is established, assuming a gray opacity:
\begin{equation}
\tau=\kappa/g P,
\label{eq:tau_P}
\end{equation}
where $\kappa$ is the gray opacity and g is the gravitational acceleration.
Therefore decreasing the surface gravity is expected to mimic the effects of
increasing metallicity (which is logarithmically related to the opacity) on the
optical depth, up to some degree. It is more convenient to combine the
metallicity and log(g) parameters with a linear relation and introduce a
modified $\beta$-factor \citep{molliere_model_2015} as follow:
\begin{equation}
\beta=\log(g)-c_\beta [Fe/H],
\label{eq:beta}
\end{equation}
where $c_\beta$ is a constant and represents relative importance of log(g) over
metallicity (for a detailed description of the $\beta$-factor see
\hyperref[sec:beta]{appendix} \;\ref{sec:beta}). Hence a decreasing $\beta$-factor lowers (C/O)\textsubscript{tr} ratio for \emph{Class-I} planets. \hyperref[fig:fig8]{Figure}\;\ref{fig:fig8} illustrates the calculated
(C/O)\textsubscript{tr} ratios for all models and the described trend is evident for the cold planets.

\begin{figure*} \includegraphics[width=\textwidth]{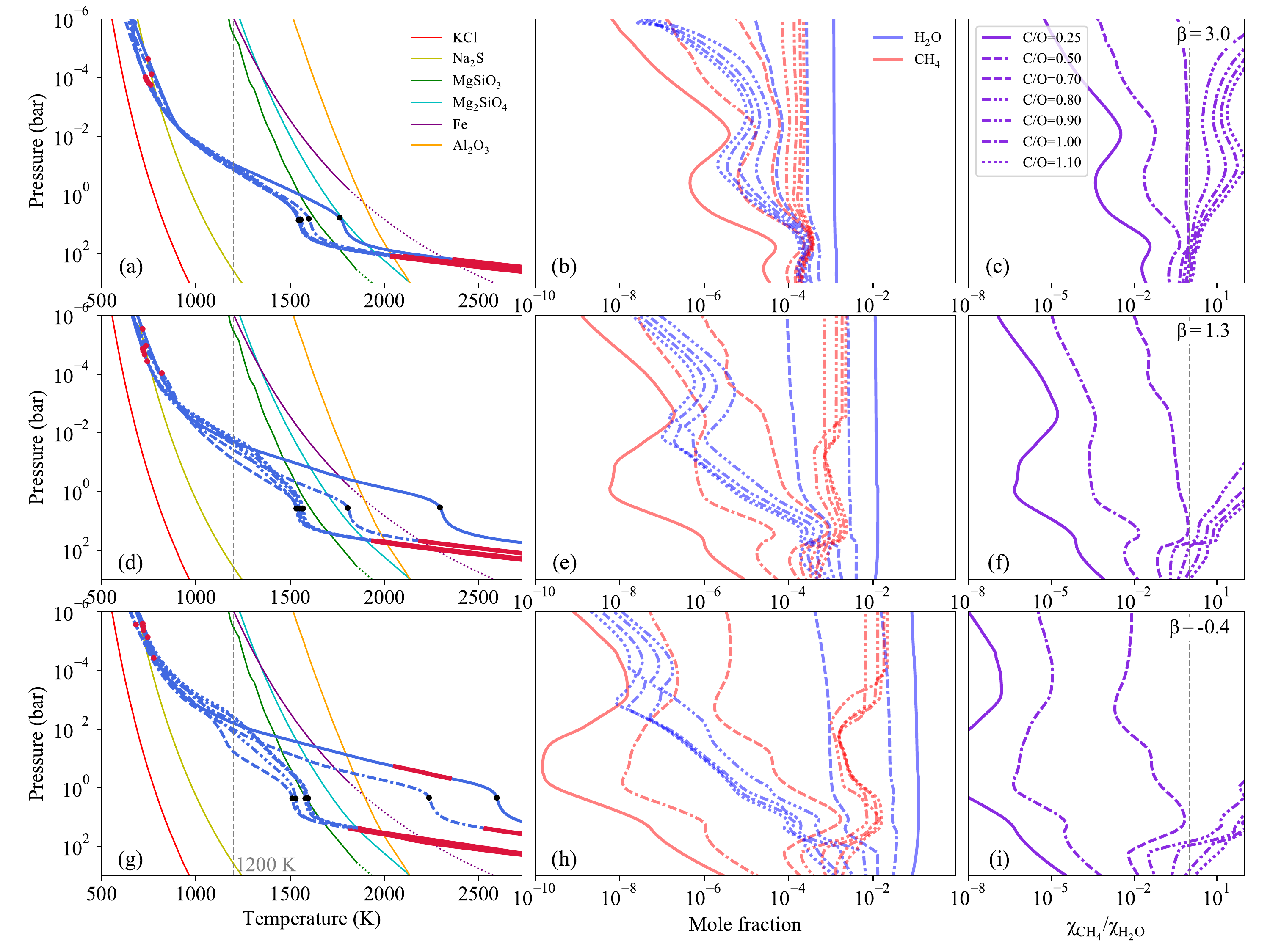} \caption{ \revision{Decreasing the $\beta$-factor increases the transition C/O ratio of Class-II planets. (a) TP structures for a planet with T\textsubscript{eff}$=$1200\;K (gray dashed line), log(g)=3.00, [Fe/H]=0.00 ($\beta$=3.0) orbiting an F5 star at different C/O ratios (blue lines); shaded red lines are convective regions; red and black dots represent 0.1\% and 99\% absorption levels of stellar flux respectively, i.e. the approximate location of the photosphere; Condensation curves are shown for solar metallicity. (b) \ce{H2O} and \ce{CH4} mole fractions at different C/O ratios. (c) \ce{CH4} to \ce{H2O} mole fraction ratios at different C/O ratios; a mole fraction ratio of unity (gray dashed line) at photospheric levels roughly coincides with the transition C/O ratio. The mole fraction ratios for C/O=0.7 model is about unity which suggests the transition C/O ratio should be close to this value. (d, e, f) Similar to (a, b, c) but with [Fe/H]=1.00 ($\beta$=1.3). (f) \ce{CH4} to \ce{H2O} mole fraction ratios decreased which hint at a higher transition C/O ratio compared with (c). (g, h, i) Similar to (a, b, c) but with [Fe/H]=2.00 ($\beta$=-0.4). (i) Mole fraction ratios continue to decrease and so do their associated transition C/O ratios.}\label{fig:fig_TPs}} \end{figure*}

The (C/O)\textsubscript{tr} ratios for \emph{Class-II} planets (with intermediate temperatures), however, demonstrate a completely different trend with respect to the \emph{Class-I}, where (C/O)\textsubscript{tr} ratios increase with higher metallicity and lower log(g), i.e. lower $\beta$-factor,
at any given temperature. As briefly discussed, this is mainly due to the presence of oxygen-bearing condensates. Higher [Fe/H] and lower log(g) pull the photosphere toward lower pressures while keeping the corresponding temperatures at the photospheric level almost the same. This lower pressure environment enhances the partial evaporation of the condensates, such as \ce{MgSiO3(c)},
\ce{MgSiO3(L)}, \ce{Mg2SiO4(c)}, \ce{Mg2SiO4(L)}, \ce{Fe2O3(c)} and \ce{Fe2SiO4(c)}, which results in a decreased \ce{CH4} \revision{but increased \ce{H2O} abundances} at the photospheric levels. Therefore the transition to a methane-dominated spectra happens at higher C/O ratios, \revision{see \hyperref[fig:fig_TPs]{Figure}\;\ref{fig:fig_TPs}}. \revision{The effect of cloud opacity and non-equilibrium chemistry on this trend yet remain to be investigated.}

The mentioned role of \ce{CO2} formation in \emph{Class-I} and partial evaporation of condensates in \emph{Class-II} are not Class-specific and both mechanisms are in action at the boundary of these two Classes and hence influence the spectral appearance. Moreover, the temperature at which this boundary occurs, i.e. the (C/O)\textsubscript{tr} minima in
\hyperref[fig:fig6]{Figures}\;\ref{fig:fig6} and\;\ref{fig:fig8}, depends on the $\beta$-factor with the \emph{Class-I}-to-\emph{Class-II} transition happening at hotter planets for higher $\beta$-factors, see \hyperref[fig:fig7]{Figure}\;\ref{fig:fig7}.

In \emph{Class-III}, (C/O)\textsubscript{tr} ratios show no substantial dependency on metallicity and surface gravity, but at higher temperatures, i.e. \emph{Class-IV}, HCN captures most of the carbon atoms in the upper atmosphere and imposes a significant depletion of remaining \ce{CH4} at high temperatures.
However, this \ce{CH4} depletion increases the transition carbon-to-oxygen values only slightly. As the photosphere rises to lower pressure at higher metallicities, water and methane abundances and the TP structure are also consistently moved to the lower pressures, and thus the contributions of water and methane features in the spectra remains alike. 

At T$>$2500\;K \ce{H2O} also starts to dissociate at low-pressure levels and in
turn makes the oxygen atoms available to other stable molecules under these
conditions. This mostly occurs at high metallicities and low C/O ratios and
appears in the spectra when log(g) is adequately high. Altogether we should
expect a mixed dependency of the transition on log(g) and metallicity in
\emph{Class-IV}. \hyperref[tab:tab1]{Table}\;\ref{tab:tab1} provides a summary
of (C/O)\textsubscript{tr} dependency on the model parameters.

\begin{table*} \centering \begin{tabular}{p{3cm}p{3cm}p{3cm}p{5cm}} Temperature
    (K) & Influencing & (C/O)\textsubscript{tr} & Dominant cause  \\ (Classes)
    & parameter &  & of dependency \\ \hline \hline $<600$-$1100$\;K & Lower
    log(g) & Decreases & Formation of \ce{CO2} \\ (Class-I) & Higher [Fe/H] &
    Decreases & at photosphere \\ \hline $600$-$1100$ to $\sim$$1650$\;k &
    Lower log(g) & Increases & Evaporation of condensates  \\ (Class-II)&
    Higher [Fe/H] & Increases & at photosphere \\ \hline $\sim$$1650$ to
    $\sim$$2200$\;K & Lower log(g) & Almost invariant & Lack of  \\ (Class-III)
    & Higher [Fe/H] & Almost invariant & condensates \\ \hline $>\sim$$2200$\;K
    & Lower log(g) & Increases & Lifting up the photosphere to lower pressure
    and dominance of HCN; \\ (Class-IV) & Higher [Fe/H] & Almost invariant &
Water dissociation at lower pressures \\ \hline \hline \end{tabular} \caption{A
summary table on the effect of log(g) and FEH} \label{tab:tab1} \end{table*}

\subsection{Impact of stellar type} \label{subsec:classes_stellar}

In addition to the surface gravity and metallicity of the planets, the spectral type of their host star also affects the (C/O)\textsubscript{tr} ratios and hence the boundary of different classes.

\begin{figure*} \includegraphics[width=\textwidth]{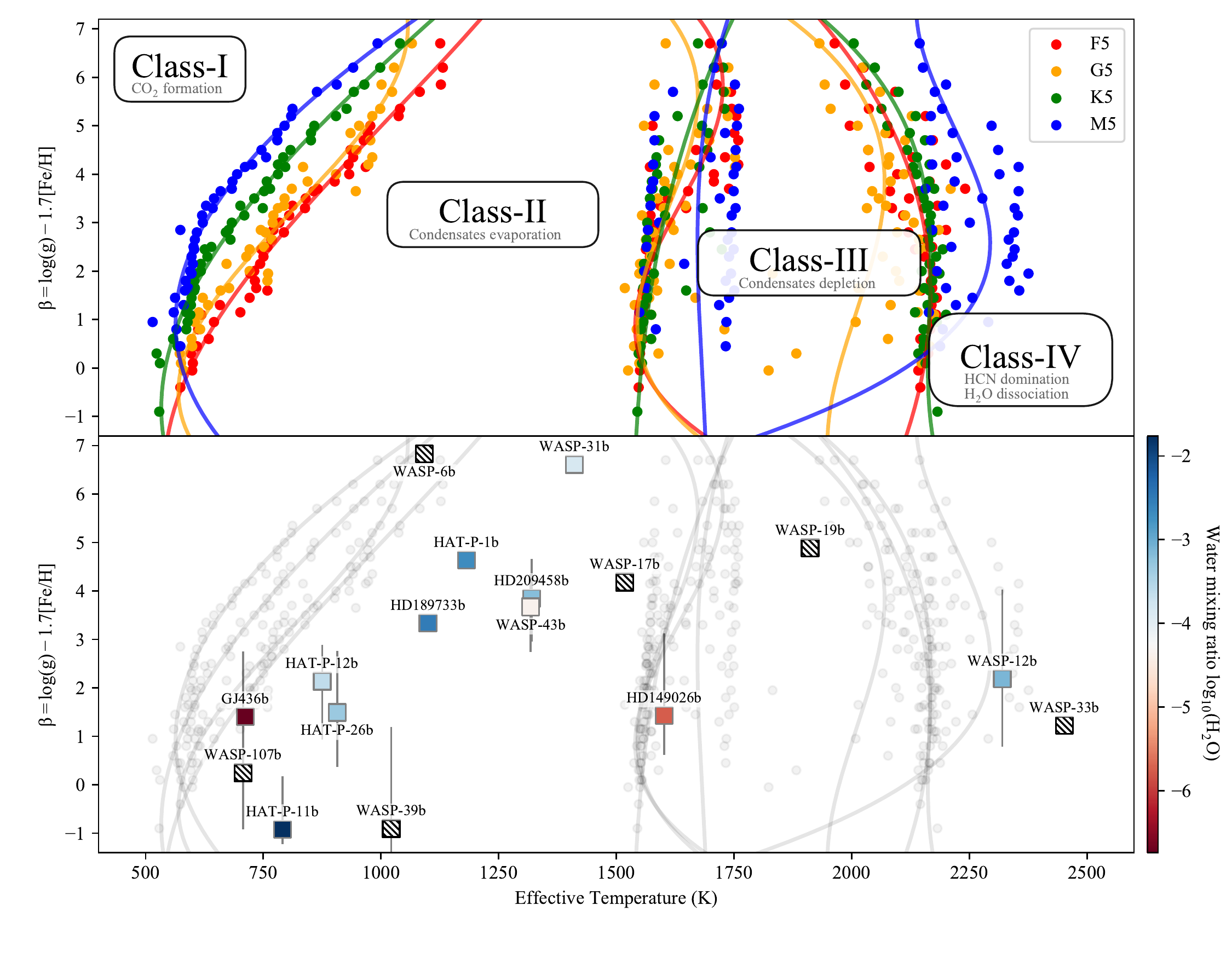}
    \caption{Top) $\beta$-T\textsubscript{eff} diagram: Mapping the effect of stellar type in four atmospheric classes. Bottom) the position of observed exoplanets with estimated metallicity and $\beta$-factor. Observational points are color-coded by the retrieved water abundances from \citet{tsiaras_population_2018}. Planets without estimation of their water content are shown by hatches. The dominant chemical mechanism at the photospheric levels of the irradiated planets vary with their effective temperature, and hence influences their spectral appearance. \revision{Note that there has been no estimation of metallicity and $\beta$-factor for Class-I planets, likely due to the presence of clouds and muted spectral features, e.g. GJ 1214b's flat transmission spectra \citep{berta_flat_2012,kreidberg_clouds_2014}.} \label{fig:fig7}} \end{figure*}

We found the boundary of \emph{Class-I} and \emph{Class-II} planets by locating the minimum (C/O)\textsubscript{tr}, for \emph{Class-II} to \emph{Class-III} by setting (C/O)\textsubscript{tr}$=$$0.9$ and likewise for \emph{Class-III} to \emph{Class-IV} by estimating the temperatures at which
(C/O)\textsubscript{tr}$=$$1.0$. We then estimated the $\beta$-factor by minimizing the scatter (see \hyperref[sec:beta]{appendix} \;\ref{sec:beta}) and separate the boundaries according to their stellar types. We found $c_{\beta}$$\sim$$1.7$ which is an indication of metallicity to be more
influential than surface gravity for the classification. \hyperref[fig:fig7]{Figure}\;\ref{fig:fig7} maps the boundaries of these four classes through a $\beta$-T\textsubscript{eff} diagram.

\begin{figure*} \includegraphics[width=\textwidth]{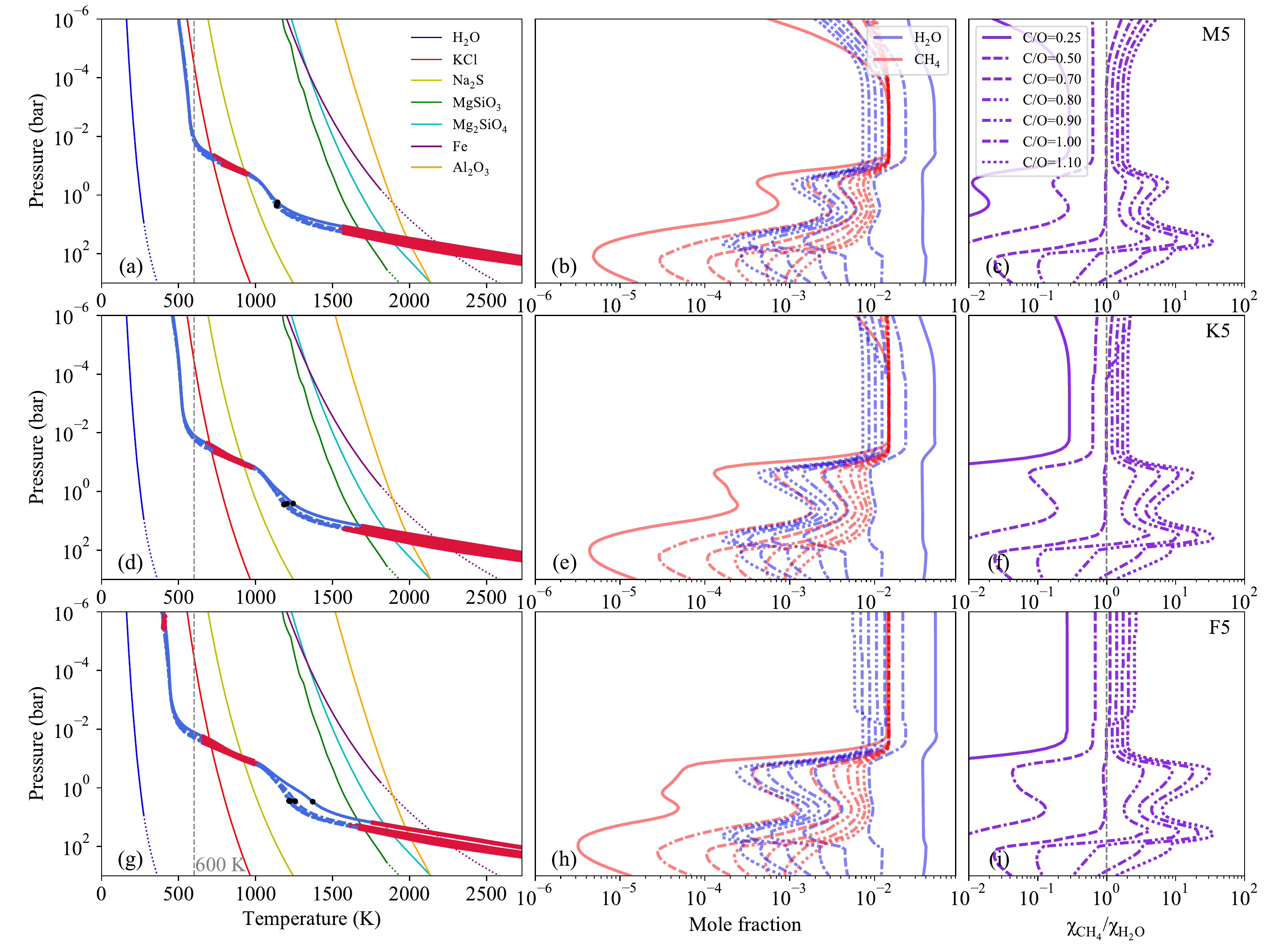} \caption{ \revision{Hotter host stars increase the transition C/O ratios in Class-I planets. (a) TP structures for a planet with T\textsubscript{eff}=600\;K (gray dashed line), log(g)=3.00, [Fe/H]=1.50 orbiting an M5 star at different C/O ratios (blue lines); shaded red lines are convective regions; red and black dots represent 0.1\% and 99\% absorption levels of stellar flux respectively; Condensation curves are shown for solar metallicity. (b) \ce{H2O} and \ce{CH4} mole fraction. (c) \ce{CH4} to \ce{H2O} mole fraction ratios; a mole fraction ratio of unity at photospheric levels roughly coincides with the transition C/O ratio which is about 0.7 in this case. (d, e, f) Similar to (a, b, c) but for K5 star. (e) \ce{CH4} abundance decreased while \ce{H2O} increased relative to M5 case; resulting in a higher transition C/O ratio. (g, h, i) Similar to (a, b, c) but for F5 host star. (h) \ce{CH4} abundance decreases while \ce{H2O} increases relative to K5 case; resulting in a slightly higher transition C/O ratio. Note that the $\beta$-factor is more dominant than the stellar type in its effect on the transition C/O ratio.}\label{fig:fig_stellartype}} \end{figure*}

The boundary between \emph{Class-I} and \emph{Class-II} is influenced by the stellar spectral type with the earlier types moving the boundary toward hotter planets. The effect is less pronounced at the lower $\beta$-factors which results in a cut-off temperature at around 550\;K where all spectral types appear to have similar transition temperature from \emph{Class-I} to \emph{Class-II} and no dependency to the stellar spectral type. The temperature at which the transition occurs can be estimated by \hyperref[eq:boundary] {Equation}\;\ref{eq:boundary}.
\begin{equation}
\label{eq:boundary}
    T_{Class I-Class II}
=
\Big(\sum_{i=0,2}
a_i(T_s)\beta^i\Big)
\end{equation}
where $a_0 = 535 + 0.005T_s$, $a_1 = -84 + 0.002T_s$, $a_2 = 20 - 0.003T_s$, $T_s$ is the host star's temperature in Kelvin and $\beta$ can be calculated through \hyperref[eq:beta]{Equation}\;\ref{eq:beta} by setting $c_{\beta}$=$1.7$. At any given $\beta$-factor hotter host stars make the boundary occur at hotter planets, mostly due to their ability to heat the atmosphere of \emph{Class-I} planets more efficiently than the late types. \revision{This leads to decreasing \ce{CH4} and increasing \ce{H2O} abundances which leads to higher transition C/O ratios at the photospheric levels. Interestingly, the transition C/O ratios are more sensitive to the $\beta$-factor than to the stellar type, as demonstrated in \hyperref[fig:fig_stellartype] {Figure}\;\ref{fig:fig_stellartype}.} Higher (C/O)\textsubscript{tr} values for earlier types results in pushing the \emph{Class-I}/\emph{Class-II} transition to higher planetary effective temperatures.

Although a slight correlation between \emph{Class-II}/\emph{Class-III} transition and the $\beta$-factor can be found, a specific effect of stellar type is difficult to deduce due to our coarse temperature step size of 200\;K. Nevertheless, the transition moves slightly to hotter planets as the $\beta$-factor increases because the photosphere moves to higher pressures in the atmosphere under these conditions, i.e. the surface gravity increases or metallicity decreases.

The stellar type has a somewhat different effect on the
\emph{Class-III}/\emph{Class-IV} transition relative to
\emph{Class-I}/\emph{Class-II}: the earlier types move the boundary to colder planets as a consequence of their efficient destruction of \ce{CH4} and \ce{H2O} by thermal dissociation.

If these trends continue to higher $\beta$-factors, all transitions converge to one transition region at T\textsubscript{eff}$\sim$1750K at $\beta$$\sim$10,
where the transition occurs from \emph{Class-I} to \emph{Class-IV} directly. Observing the atmosphere of such planets with extreme surface gravity and metallicity is very difficult, if not impossible, due to their small scale height. The question arises if such planets exist at all.

\begin{table*} \centering
    \begin{tabular}{p{1.75cm}p{1cm}p{1.25cm}p{1.5cm}p{1.5cm}p{1.5cm}p{6cm}}
        Planet & $T_{eq} (K)$ & log(g) & [Fe/H] & $\beta$ & C/O &  Ref. \\
        \hline \hline CoRoT-2b & 1393 & $3.60_{-0.03}^{+0.03}$ & -- & -- &
        $0.8^{+0.5}_{-0.5}$ & \citet{madhusudhan_c/o_2012} \\ \hline GJ436b &
        712 & $3.09_{-0.05}^{+0.05}$ & 1.0 & 1.41 & $1.0^{?}_{-0.5}$ &
        \citet{madhusudhan_high_2011,line_systematic_2014} \\ \hline HAT-P-1b &
        1182 & $2.90_{-0.02}^{+0.02}$ & -1.0 & 4.64 & 0.15 &
        \citet{goyal_library_2017} \\ \hline HAT-P-11b & 791 &
        $3.07_{-0.06}^{+0.06}$ & $2.30^{+0.18}_{-0.65}$ & $-0.92^{1.1}_{-0.3}$
        & -- & \citet{fraine_water_2014,wakeford_hat-p-26b:_2017} \\ \hline
        HAT-P-12b & 875 & $2.77_{-0.04}^{+0.04}$ & $0.37^{+0.7}_{-0.45}$ &
        $2.1^{+0.75}_{-1.2}$ & $0.57^{+0.06}_{-0.07}$ &
        \citet{goyal_library_2017}; Yan et al in prep. \\ \hline HAT-P-26b &
        907 & $2.68_{-0.09}^{+0.10}$ & $0.70^{+0.66}_{-0.74}$ &
        $1.48^{+1.27}_{-1.13}$ & -- & \citet{wakeford_hat-p-26b:_2017} \\
        \hline HD149026b & 1602 & $3.25_{-0.05}^{+0.05}$ & $1.0^{+0.48}_{-1.0}$
        & $1.43^{+1.7}_{-0.8}$ & $0.55^{+0.45}_{-0.1}$ &
        \citet{fortney_atmosphere_2006,line_systematic_2014,zhang_phase_2018}
        \\ \hline HD189733b & 1100 & $3.37_{-0.03}^{+0.03}$ & $0.0$ & $3.34$ &
        $0.56^{+0.34}_{-0.09}$ &
        \citet{line_systematic_2014,goyal_library_2017} \\ \hline HD209458b &
        1320 & $2.97_{-0.02}^{+0.02}$ & $-0.52^{+0.52}_{-0.48}$ &
        $3.85^{+0.81}_{-0.89}$ & $0.56^{+0.44}_{?}$ &
        \citet{line_systematic_2014,line_no_2016,goyal_library_2017} \\ \hline
        WASP-6b & 1092 & $2.94_{-0.04}^{+0.04}$ & -2.30 & 6.83 & 0.15 &
        \citet{goyal_library_2017} \\ \hline WASP-12b & 2320 &
        $3.08_{-0.05}^{+0.06}$ & $0.48^{+0.82}_{-1.08}$ &
        $2.19^{+1.83}_{-1.40}$ & $0.51^{+0.19}_{-0.21}$ &
        \citet{stevenson_transmission_2014,line_systematic_2014,kreidberg_detection_2015,wakeford_complete_2017,goyal_library_2017}
        \\ \hline WASP-14b & 1719 & $4.06_{-0.06}^{+0.06}$ & -- & -- &
        $0.75^{+0.65}_{-0.65}$ & \citet{madhusudhan_c/o_2012} \\ \hline
        WASP-17b & 1518 & $2.50_{-0.05}^{+0.05}$ & -1.0 & 4.18 & 0.15 &
        \citet{goyal_library_2017} \\ \hline WASP-19b & 1912 &
        $3.17_{-0.02}^{+0.02}$ & -1.0 & 4.89 & $0.40^{+0.70}_{-0.25}$ &
        \citet{madhusudhan_c/o_2012,line_systematic_2014,goyal_library_2017} \\
        \hline WASP-31b & 1411 & $2.72_{-0.04}^{+0.04}$ & -2.30 & 6.61 & 0.35 &
        \citet{goyal_library_2017} \\ \hline WASP-33b & 2452 &
        $3.33_{-0.02}^{+0.02}$ & 1.48 & 1.23 & $0.75^{+0.25}_{-0.25}$ &
        \citet{madhusudhan_c/o_2012,zhang_phase_2018} \\ \hline WASP-39b & 1022
        & $2.65_{-0.06}^{+0.05}$ & $2.08^{+0.40}_{-1.23}$ &
        $-0.90^{+2.10}_{-0.68}$ & 0.56 &
        \citet{goyal_library_2017,wakeford_complete_2017} \\ \hline WASP-43b &
        1318 & $3.69_{-0.02}^{+0.02}$ & $0.00^{+0.54}_{-0.40}$ &
        $3.67^{+0.68}_{-0.92}$ & $1.0^{?}_{-0.87}$ &
        \citet{line_systematic_2014,kataria_atmospheric_2015,wakeford_complete_2017}
        \\ \hline WASP-107b & 707 & $2.55_{-0.04}^{+0.04}$ &
        $1.34^{+0.69}_{-1.47}$ & $0.25^{+2.50}_{-1.17}$ &
        $0.02^{+1.58}_{-0.01}$ &
        \citet{anderson_discoveries_2017,kreidberg_water_2018} \\ \hline XO-1b
        & 1168 & $3.21_{-0.03}^{+0.03}$ & -- & -- & $0.80^{+0.20}_{-0.20}$ &
        \citet{madhusudhan_c/o_2012} \\ \hline \hline

\end{tabular} \caption{A summary of estimated metallicity and carbon to oxygen
    ratio for 20 exoplanets\tablenotemark{**}} \label{tab:tab2}
    \tablenotetext{\tiny **}{  One should be cautious about the [Fe/H] and C/O
    values summarized in this Table due to different data reduction,
    assumptions in the forward simulations or retrieval techniques. Here we
    report them only to provide a qualitative picture of the current status of
    our understanding regarding these parameters in the context of locating
    their positions on the planetary classification map,
    \hyperref[fig:fig7]{Figures}\;\ref{fig:fig7} and\;\ref{fig:fig8}.}
    \end{table*}

The bottom panel of \hyperref[fig:fig7]{Figure}\;\ref{fig:fig7} illustrates the location of some of the observed planets for which metallicities have been estimated. The data
are summarized in \hyperref[tab:tab2]{Table}\;\ref{tab:tab2}. 
The metallicities are not always well constrained and, in some cases, are only reported to be consistent with the observations. Clearly, a coherent analysis of available data and additional observations are needed to draw any conclusions. 
Nevertheless, the Figure shows that none of the observed planets are characterized as \emph{Class-I}, mostly due to their smaller scale heights and possible cloud coverage in the case of transmission spectroscopy \revision{(e.g. GJ 1214b's flat transmission spectra \citep{berta_flat_2012,kreidberg_clouds_2014})}, and lower emergent flux in the case of emission spectroscopy. Most of the observed planets in fact belong to the \emph{Class-II}; this is mainly a consequence of the higher number of detected planets in this temperature range.

\revision{Since emission spectra probe deeper than transmission spectra into the atmosphere one might ask how the boundaries of four classes would change if the analysis was based on the emission spectra. Applying the spectral decomposition technique (\hyperref[sec:decomposition]{Section}\;\ref{sec:decomposition}) on emission spectra is technically challenging, mostly due to the necessity of making different templates for each model based on their exact TP structure. However, \hyperref[fig:fig3]{Figure}\;\ref{fig:fig3} could provide some insight. The pressure of a given photospheric level increases as $\beta$-factor increases. Therefore if we employ emission spectra instead, we would probe higher pressures for a given photospheric level. Consequently, a lower $\beta$-factor would be required to keep a given photospheric level at the same pressure for both transmission and emission spectra. Thus a slight shift of the $\beta$-T\textsubscript{eff} diagram (\hyperref[fig:fig7]{Figure}\;\ref{fig:fig7}) toward lower $\beta$ values is expected if our analysis was based on the emission spectra.}

\subsection{A Parameter Space for \ce{CH4}} \label{subsec:CH4}

The lack of a robust methane detection in the spectra of irradiated exoplanets (see e.g. \citealt{madhusudhan_high_2011}) immediately rises a question: \emph{Does there exist a parameter space preferential for the detection of \ce{CH4}}?

\cite{espinoza_metal_2017} studied C/O ratios of 50 cold planets with
T\textsubscript{eff}$<$1000\;K. They found C/O$>$1 to be highly unlikely for
these planets and concluded that this is likely to be a universal outcome for gas giants. By extending this conclusion to hot planets one could expect the spectra of \emph{Class-IV} planets to be always water-dominated with no room for a methane dominated spectrum. This reduces the probability of \ce{CH4} detection at high temperatures, i.e T\textsubscript{eff}$>$1500\;K, significantly.

However, \hyperref[fig:fig8]{Figure}\;\ref{fig:fig8} reveals a region between 800 and 1500\;K with C/O$>$0.7 where the \ce{CH4} molecule is predicted to always be the dominant spectral feature. We therefore call this region
\emph{the Methane Valley} \revision{and anticipate higher probability of \ce{CH4} detection over \ce{H2O} spectral features when exploring this parameter space}.

The lack of such a detection would likely to be an indication that \revision{cloud-free models are incapable of capturing exo-atmosphere characteristics within that parameter space}. Hence, studying the methane valley could potentially provide a suitable road-map to study departures from thermo-chemical equilibrium, departures from radiative-convective hydrostatic
equilibrium, or the effect of clouds on the presence of methane in planetary atmospheres. \revision{Mapping the observed planets with estimated C/O on \hyperref[fig:fig8]{Figures}\;\ref{fig:fig8}, tentatively suggest such departure from cloud-free equilibrium chemistry conditions where the observed planets within or close to this region (i.e. WASP-43b, XO-1b, HD 189733b, and HAT-p-12b) all have shown significant water features in their transmission spectra. More precise C/O measurements are needed to observationally constrain the Methane Valley properties.}

\subsection{Color-Temperature Diagrams} \label{subsec:color_diagram}

\begin{figure*} \gridline{\fig{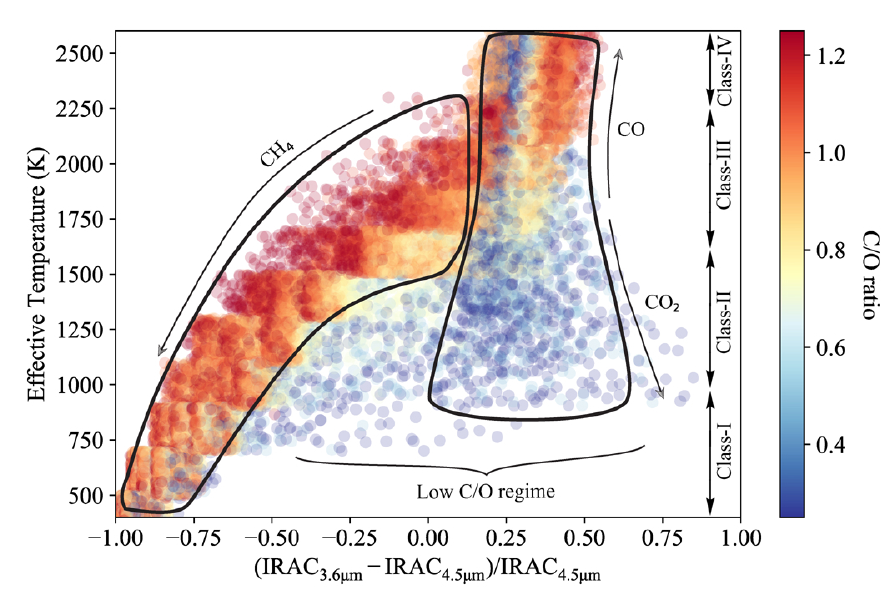}{0.5\textwidth}{} \fig{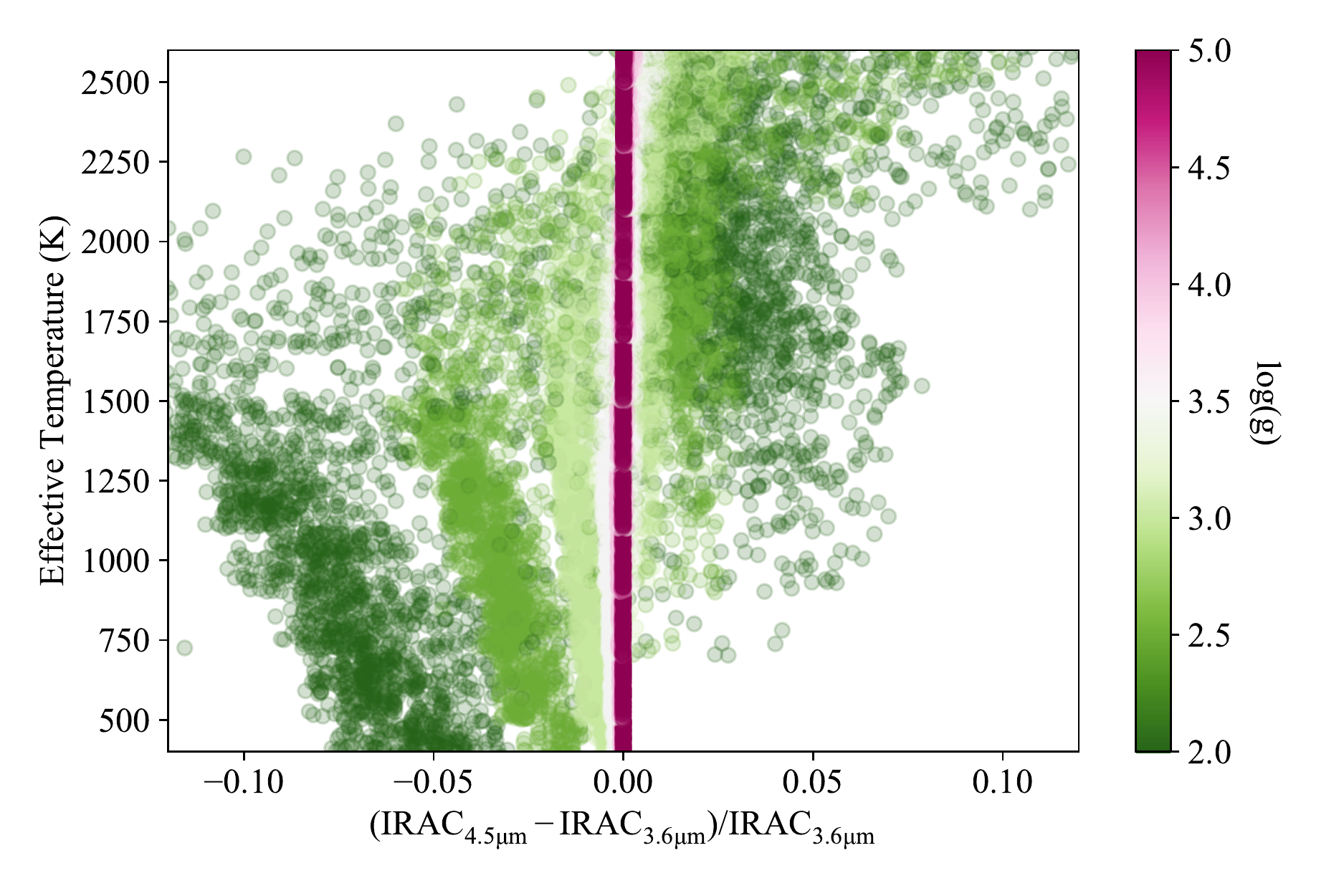}{0.5\textwidth}{} } \caption{Synthetic IRAC color-temperature diagrams for cloud free atmospheres under equilibrium chemistry condition. Left) Color diagram based on emission spectroscopy, i.e. IRAC data describes the secondary eclipse depth at $\lambda$ $(\mu m)$. Right) The same for transmission technique, i.e. IRAC data describes the transit depth at $\lambda$ $(\mu m$). \label{fig:color_diagram}}
\end{figure*}

Color-diagrams have been used as a method of characterization of self-luminous objects for more than a century, see e.g. \citet{rosenberg_uber_1910} for stars, \citet{tsuji_transition_2003} for cool dwarfs and brown dwarfs, and \citet{bonnefoy_library_2014,keppler_discovery_2018,batalha_color_2018} for directly imaged planets. \revision{\citet{triaud_colourmagnitude_2014} and \citet{triaud_colourmagnitude_2014-1} studied color-magnitude diagrams of known transiting exoplanets. Their investigations, therefore, were limited to the systems with known parallaxes. They compared irradiated planets with very low mass stars and field brown dwarfs and concluded that further measurements are required to confirm or reject whether irradiated gas giants form their own sequence on the color-magnitude diagrams.}

Inspired by these works, we investigated the possibility of introducing a color-diagram for the characterization of irradiated planets, \revision{using their effective temperature instead of their absolute magnitude. The effective temperature can be used as a proxy for the luminosity/absolute magnitude, because the reference radius assumed to be constant, see \hyperref[sec:results]{Section}\;\ref{sec:results}.} To take an even more practical approach we \revision{chose a normalized color parameter based} on the Spitzer's Infrared Array Camera (IRAC) \citep{fazio_infrared_2004,allen_infrared_2004} as a commonly used photometer for the observation of exoplanets (see e.g. \citet{charbonneau_detection_2005,deming_spitzer_2007,todorov_spitzer_2009,sing_continuum_2016}). The IRAC photometric channels 1 and 2 are centered at 3.6 and 4.5\;$\mu$m, respectively. Channel 1 (3.6\;$\mu$m) is more suited to study \ce{CH4}/\ce{H2O} spectral features while channel 2 (4.5\;$\mu$m) is more sensitive to
CO/\ce{CO2} features (see e.g.
\citealt{swain_molecular_2009,desert_search_2009,swain_water_2009}). Depending on the type of spectroscopy, i.e. transmission or emission, the ratio of the transit depth or the ratio of the secondary eclipse depth at these channels could potentially provide an information regarding the relative presence of these molecules in the atmosphere of a planet. For transmission spectroscopy, we define this ratio as:
\begin{equation}
    \label{eq:IRAC_trans} 
R_{tr}=(IRAC_{4.5\;\mu m}
-IRAC_{3.6\;\mu
m})/IRAC_{3.6\;\mu m}
\end{equation}
where $IRAC_{\lambda}$ is the transition depth observed at wavelength $\lambda$ ($\mu$m) channel. In transmission spectroscopy absorption features appear as positive signals in the unit of transit depth. This is not the case for emission spectroscopy where absorption features are negative signals with respect to a blackbody curve. As a result, we rearrange the terms and define this ratio of channels for emission spectroscopy as follows:
\begin{equation}
\label{eq:IRAC_emission}
    R_{em}=(IRAC_{3.6\;\mu
m}-IRAC_{4.5\;\mu m})
/IRAC_{4.5\;\mu m}
\end{equation}
where $IRAC_{\lambda}$ is the secondary eclipse depth observed at wavelength
$\lambda$ ($\mu$m) channel. By applying IRAC's spectral response curves on our
56,448 synthetic spectra we estimated these ratios for the two spectroscopy
methods. \hyperref[fig:color_diagram]{Figure}\;\ref{fig:color_diagram} shows
the IRAC synthetic color-temperature diagrams for cloud-free atmospheres under
equilibrium chemistry conditions. 

The general shape of the ``emission" color-temperature diagram
(\hyperref[fig:color_diagram]{Figure}\;\ref{fig:color_diagram}, left panel) is very similar to that of self-luminous dwarfs and directly imaged planets, with two distinct populations at the first glance. One population is associated with colder planets and negative $R_{em}$ values, which means the planetary emission
flux at 4.5\;$\mu$m is stronger than at 3.6\;$\mu$m. This in turn indicates that the absorption at 3.6\;$\mu$m is stronger and, hence, suggests the presence of strong \ce{CH4} over CO/\ce{CO2} features. In contrast, the second hotter population has positive $R_{em}$ values and, therefore, indicates pronounced CO/\ce{CO2} spectral features. A typical uncertainty of observed $R_{em}$ by IRAC is on the order of $\sim$0.25 and, consequently, the two populations should be distinguishable. Any deviation from these two populations could be a consequence of eddy diffusion, the presence of clouds or sub-solar C/O in the visible atmosphere of exoplanets. The case of low C/O is evident for the planets with an effective temperature between 750 and 1750\;K, and C/O$<$0.4 in the left panel of \hyperref[fig:color_diagram]{Figure}\;\ref{fig:color_diagram}, which shows significant scatter.

From the transmission color-temperature diagram (see
\hyperref[fig:color_diagram]{Figure}\;\ref{fig:color_diagram}, right panel), a
new pattern emerges, but the two populations are still distinguishable: the population of colder planets with negative $R_{em}$ values (with stronger \ce{CH4} spectral features) and the population of hotter planets with positive $R_{tr}$ values (with stronger CO/\ce{CO2} features). \revision{The diagram is color-coded by surface gravity values, because the amplitude of transmission spectral features is strongly correlated with this parameter, i.e. higher log(g) results in smaller features (see e.g. \citet{lecavelier_des_etangs_rayleigh_2008}).} A typical uncertainty of observed $R_{tr}$ by IRAC is on the order of $\sim$0.05. However only planets with log(g)$\leq$$3.0$ result in a $R_{tr}$$\geq$$0.05$. Therefore, any significant deviation from $R_{em}$$\sim$$0$ for planets with log(g)$\geq$$3.0$ should be due to the effects of non-equilibrium chemistry or clouds on their transmission spectra. This could be used as a diagnostic tool to indicate such planets as suggested by \citet{baxter_comprehensive_2018}.

\section{SUMMARY AND CONCLUSION} \label{sec:conclusion}

In this paper we have studied the dominant chemistry in the photosphere of irradiated gaseous exoplanets by calculating a large grid of self-consistent cloud-free atmospheric models, \revision{as a preparatory step toward a framework for an observationally driven classification scheme.}

\emph{The Spectral Decomposition Technique} enabled us to quantitatively estimate the contribution of \ce{H2O} and \ce{CH4} in the synthetic transmission spectra and hence we were able to find the transition C/O ratios at which the water-dominated spectrum flips to a methane-dominated one. We find that C/O$<$1 is not a global indicator of water-dominated spectra, and C/O$\geq$1 is not a ubiquitous indication of methane-dominance, see e.g. \hyperref[fig:fig8]{Figure}\;\ref{fig:fig8}. However, the separation at C/O$=$1 still provides a rough approximation for the water-methane boundary for adequately hot planets.

Mapping all the transition C/O ratios revealed four spectral populations of planets in the C/O-T\textsubscript{eff} diagram. Hence a ``four-class" classification scheme emerged for irradiated planets; spanning from cold (400\;K) to hot (2600\;K) planets. The spectra within the temperature range of 600\;K to 1100\;K, i.e. the boundary of \emph{Class-I} and \emph{Class-II}., is found to be quite diverse and a slight variation of the physical parameters, such as metallicity or surface gravity, could lead to another chemistry and hence to another spectral class. This parameter space is thus well-suited for studying the diversity of physics and chemistry of exoplanetary atmospheres. Such study potentially opens the path to the study of colder planets.

We have also predicted a region (\emph{The Methane Valley}) where methane always remains the cause of dominant spectral features, under the assumptions of this study. The temperature range to look for \ce{CH4} features spans from 800 to 1500\;K and requires C/O$\geq$$0.7$. Although \ce{CH4} is expected to be more present in the atmosphere of colder planets, the temperature range of the Methane Valley is expected to be  in favor of less cloudy and less vertically quenched atmospheres, which increases the probability of \ce{CH4} detection in turn. \ce{CH4} detection in the Methane Valley, or the lack of it, could hint the prevalence of cloud formation or non-equilibrium chemistry within this parameter space and provides a diagnostic tool to identify these conditions.

We constructed two Spitzer IRAC color-diagrams; one from the synthetic transmission and one from emission spectra. In both cases two populations of planets can be interpreted. One population highlights the planets with a stronger \ce{CH4} photometric signature at 3.6\;$\mu$m (mostly associated with \emph{Class-I}, \emph{II} and \emph{III}) and the other one shows a stronger CO/\ce{CO2} signal at 4.5\;$\mu$m (mostly associated with \emph{Class-II}, \emph{III} and \emph{IV}). Future photometric analysis could reveal whether irradiated planets follow the location of these populations on the color-maps or they would deviate from the predictions and hence mark the possibility of cloud presence or non-equilibrium chemistry in their photosphere.

\revision{As mentioned, the results of this paper is based on cloud-free equilibrium-chemistry assumptions. Including additional physics (such as non-equilibrium chemistry or clouds) or opacities (such as TiO/VO; see e.g. \citet{mancini_lower_2013},\citet{haynes_spectroscopic_2015},\citet{evans_detection_2016}, and \citet{nugroho_high-resolution_2017}) can potentially change the results. Therefore, our classification scheme might lead to biased conclusions in cases where the cloud formation or disequilibrium chemistry are expected to occur, e.g. colder planets, and should be taken only as an initial step toward an observationally driven characterization scheme for exoplanet atmospheres.}

\section{acknowledgment} \label{sec:acknowledgment} \revision{We thank the anonymous reviewer for the careful reading of our manuscript and her/his many insightful comments and suggestions.} This research has made use of the NASA Exoplanet Archive, which is operated by the California Institute of Technology, under contract with the National Aeronautics and Space Administration under the Exoplanet Exploration Program.

\appendix

\section{Problematic parameter space} \label{sec:problem_petit} 
\begin{figure}[ht!] \plotone{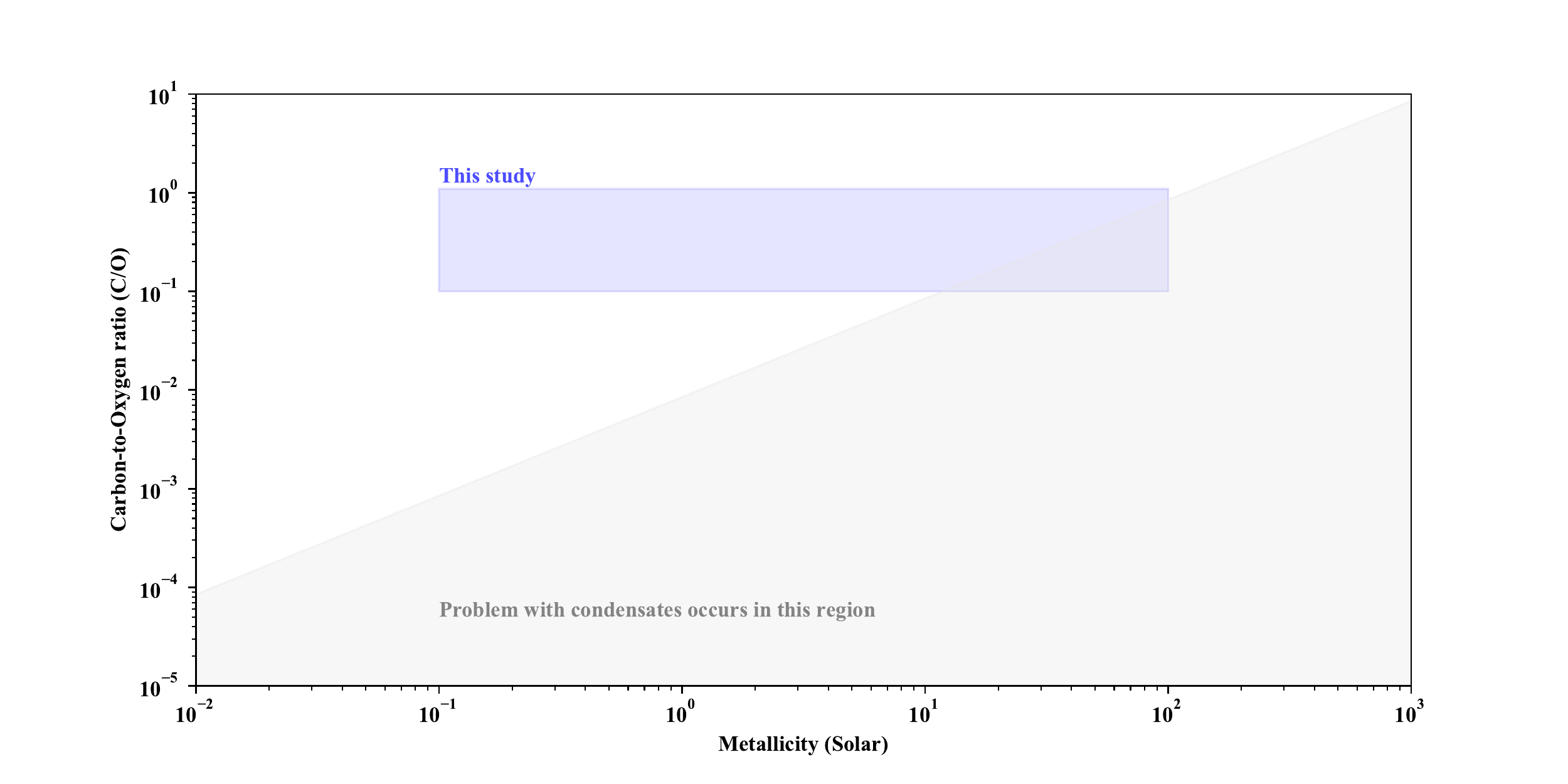} \caption{Problematic parameter
space at high metallicities and low C/Os \label{fig:fig2}} \end{figure}

With five
parameters to explore, the number of models exceeds 28,000. It then should not
be a surprise 
that some models 
demanded extra attention as a result of their poor
convergence. In particular, models with high metallicity, but low C/O turned
out to be quite problematic. By examining a wider range of metallicities and
C/O ratios the trend became more prominent, see Figure\;\ref{fig:fig2}.

This problem has been noticed before by Paul Molli\`ere and Pascal Tremblin [private communication, 2018] and it is thought to be a numerical issue with the Gibbs free energy minimization algorithm. Their investigation have shown that excluding \ce{MgSiO3(c)}, \ce{MgSiO3(L}), \ce{MgAl2O4(c)} and \ce{Fe2SiO4(c)} from the chemistry network solves the numerical problem and hence they
concluded there are possibly some linear combinations of condensates that make the abundance matrix in the Gibbs minimization problem rank-deficient. Two examples of such net reactions are:
\begin{equation}
\ce{MgAl2O4(c) + MgSiO3(c) 
<=> Mg2SiO4(c) + Al2O3(c) }
\end{equation}
\begin{equation} \ce{Fe2SiO4(c) + MgAl2O4(c) 
<=> 2FeO(c) + MgSiO3(c) + Al2O3(c)
} \end{equation}
The presence of these condensates have a prominent effect in the planetary
atmospheres as they bind oxygen from the atmosphere and change the water
content along with other compositional variations. As a result the TP structure
and resulting spectra change too. The spectral difference between the models
with and without these four condensates was explored quantitatively and more
than 10\% of our models were found to show a deviation larger than 10ppm in
their transmission spectrum (assuming Jupiter-sized planets and solar-sized
host stars). All these models had very low log(g), i.e. 2.0-2.5, in the grid.
Planets with log(g)=2.0-2.5 are not common, see
\hyperref[fig:logg]{Figure}\;\ref{fig:logg}, and also 10ppm is likely below the
expected JWST noise floor (e.g.
\citealt{ferruit_observing_2014,beichman_observations_2014,batalha_pandexo:_2017})
but nevertheless it is important to understand how removing the condensates can
improve the numerical convergence by keeping the change in the spectra to a
minimum.

We explored
the problematic models from our grid by excluding the condensates
one-by-one and found removing either of these condensates can solve the problem
of our grid: \ce{Fe2SiO4(c)}, \ce{MgAl2O4(c)}, FeO(c), \ce{Mg2SiO4(c)} and
\ce{Al2O3(C)}. The following net reaction can be a possible linear combination,
which results in a rank deficiency:
\begin{equation}
\ce{Fe2SiO4(c) + 2MgAl2O4(c) 
<=> 2FeO(c) + Mg2SiO4(c) +
2Al2O3(C) }
\end{equation}
We extended our investigation beyond our parameter space to explore this numerical issue in more detail and found that excluding \ce{Mg2SiO4(c)} can solve the problem while keeping the spectral change to a minimum. In conclusion
less than 0.7\% of our entire models have spectral change on the level of 10ppm
or above if we exclude \ce{Mg2SiO4(c)}, and thus we removed this condensate from
the problematic models to compute the grid of models.

\section{$\beta$: the scaling factor} \label{sec:beta} The relationships
between TP structures, abundances and spectra to the surface gravity and
metallicity have been reported by previous studies (see e.g.
\citealt{molliere_model_2015}). They all reported low log(g) imitates high
[Fe/H] and high log(g) emulates low metallicity effects. This can be understood
by assuming gray opacities in an atmosphere and deriving a simple relation
between the optical depth ($\tau$) and pressure (P) as noted in
\hyperref[eq:eq5]{Equation}\;\ref{eq:eq5}. \citet{molliere_model_2015}
suggested $\beta=log(g)-[Fe/H]$ as a factor to map the optical depth-pressure
in the planetary atmospheres. This linear relation is inspired by the
dependency of log(p) to the surface gravity and metallicity at any given
optical depth:
$\tau=\kappa/g P \Rightarrow P=\tau g/\kappa$
\begin{equation}
\label{eq:eq6}
\tau_{phot} \approx 1 \Rightarrow P \approx
P_{phot}=g/\kappa \Rightarrow 
\log(P_{phot}) \approx \log(g)-\log(\kappa) 
\approx \log(g) - [Fe/H] + cst 
\Rightarrow \log(P_{phot}) 
\approx \beta + cst
\end{equation}
where $P_{phot}$ is the pressure at photospheric levels. Molecular and atomic
opacities are non-gray in nature and thus the defined relation holds true only
over a narrow pressure level (or optical depth, alternatively). Hence from a
broader perspective the $\beta$-factor can be expressed as a function of
pressure or can be defined for a specific feature in the atmosphere such as the
location of convective-radiative boundary in the atmosphere and how it varies
with $\beta$.

\begin{figure*} \includegraphics[width=\textwidth]{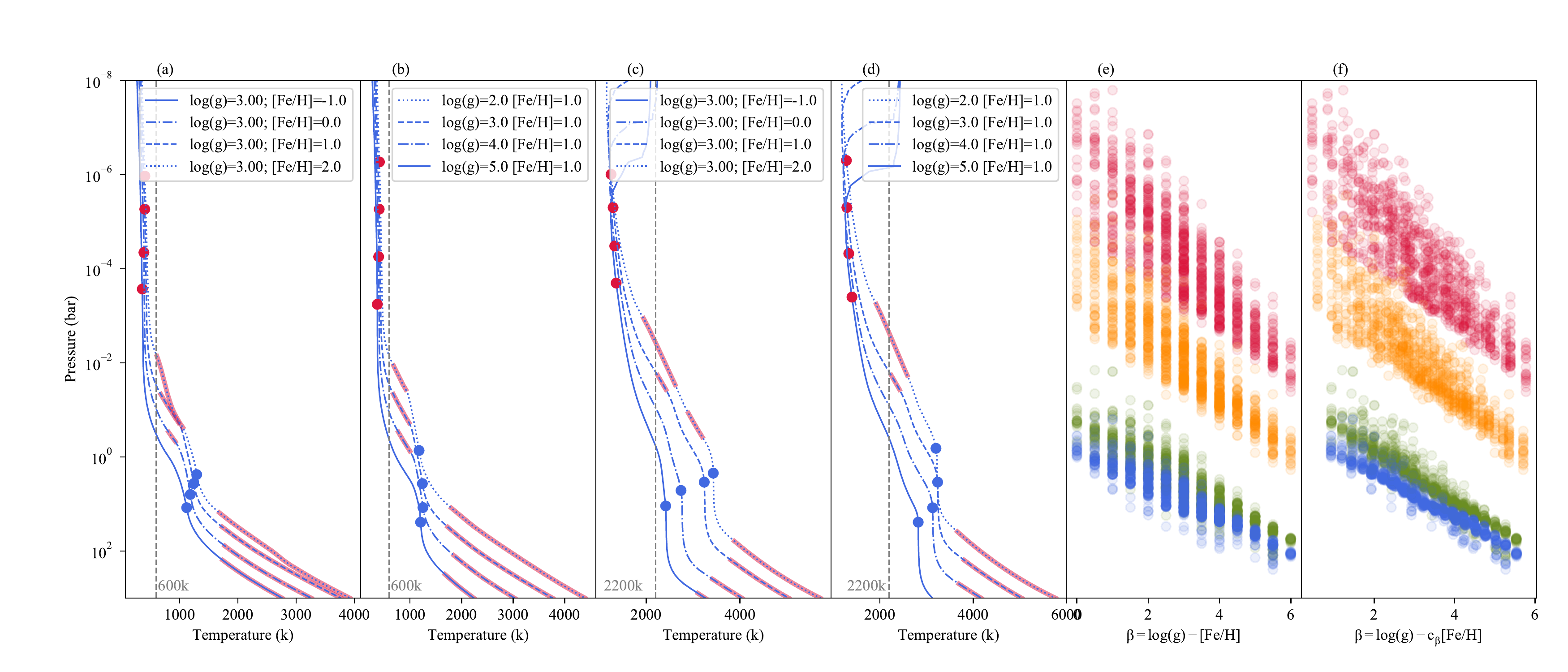}
    \caption{The effect of surface gravity and metallicity on the temperature
    structures of planetary atmospheres. \revision{(a) TP structures for planets with T\textsubscript{eff}=600\;K (gray dashed line), log(g)=3.00, C/O=0.5 orbiting an F5 star at different metallicities [Fe/H]=-1.0,0.0,1.0 and 2.0 (blue lines); Shaded red lines are convective regions; red and blue dots represent 0.1\% and 99\% absorption levels of stellar flux, respectively; (b) similar to (a) but for a fixed [Fe/H]=1.0 and varying log(g)=2.0, 3.0, 4.0, and 5.0; (c, d) similar to (a, b) but for a hot planet T\textsubscript{eff}=2200\;K.} These values are chosen to represent similar $\beta$-factor for both sets and hence TP profiles are expected to change similarly. \revision{(e) Color dots are 0.1\% (red), 5\% (orange), 95\% (green) and 99.9\% (blue) absorption levels of stellar flux drawn from 1000 randomly chosen models with their representative $\beta$-factor using $\beta=\log(g)-$[Fe/H]. (f) Finding $c_\beta$ values by minimizing the scatter of absorption levels using $\beta=\log(g)-c_\beta$[Fe/H]. See the text for more details.}
    \label{fig:fig3}} \end{figure*}

Figure\;\ref{fig:fig3}a shows the effect of increasing metallicity on the TP
structure of a cold planet with an effective temperature of 600\;K at a fixed
C/O ratio of 0.5, orbiting around a F5 star. In contrast, increasing the
surface gravity pushes the TP structure to higher pressures,
\hyperref[fig:fig3]{Figure}\;\ref{fig:fig3}b.
\hyperref[fig:fig3]{Figure}\;\ref{fig:fig3}c and d similarly demonstrate these
trends but for a hot planet with an effective temperature of 2200\;K. We locate
the pressure levels at which the stellar flux is absorbed by 0.1\% (red dots)
and 99.9\% (blue dots) on the TP structure of these models to investigate how
their photosphere vary with the beta factor. In these two cases, where the
stellar type and C/O are fixed, the location of photosphere is almost invariant to the planets’ temperatures.

1000 random models are drawn from all 28,224 simulations, regardless of their
input parameters, and their 0.1\%, 5\%, 95\% and 99.9\% stellar absorption
levels are shown in \hyperref[fig:fig3]{Figure}\;\ref{fig:fig3}e with respect
to their beta factor. At lower beta values, i.e. lower log(g) with higher
[Fe/H], the data are more scattered around a linear trend due to the pressure
dependence of the atomic and molecular opacities and the effect of C/O on the
shape of TP structure and abundances\textsc{\char13} vertical distribution. In
addition, each absorption level has slightly different dependency on the
$\beta$ factor: low pressure regions have steeper linear slopes but more scattered
in general. This suggests the relationship between the pressure levels of
spectrally active regions, and metallicity and surface gravity is possibly not
a one-to-one association. Constructing a complex function for the
$\beta$-factor is possible, however we aim to keep this modification at a
minimal level. We therefore introduce a modified relation for the beta factor
as follows:
\begin{equation}
\beta=\log(g)-c_\beta [Fe/H]
\label{eq:eq7}
\end{equation}
where $c_\beta$ is a constant and represents relative importance of log(g) over
metallicity. We define $c_\beta$ in a way to make log(g) and $c_\beta$[Fe/H]
terms comparable. Therefore, if $c_\beta$$<$1 then log(g) is influencing that
specific layer of the atmosphere more than [Fe/H], and otherwise for
$c_\beta$$>$1. When $c_\beta$=1, surface gravity and metallicity are equally
important for the region under study.

An approach to find $c_\beta$ could be to minimize the scatter at each region
of interest, for instance through $\chi^2$ minimization. By following this
approach we estimate $c_\beta$ for the $\beta$ factor at 0.1\%, 5\%, 95\% and
99.9\% stellar absorption levels to be 0.774, 0.733, 0.706, 0.534, 0.530 and
0.538, respectively. All evaluated $c_\beta$ are less than 1.0, pointing at the
surface gravity to be more influential than the metallicity on the TP
structures in the spectrally active regions, but in particular at the optically
thicker layers where $c_\beta$ is the minimum. It is also noticeable that the
deeper regimes are less influenced by other parameters such as planet’s
effective temperature, stellar type or carbon to oxygen ratio. This can be seen
in the less scattered $\beta$-factor in
\hyperref[fig:fig3]{Figure}\;\ref{fig:fig3}f for the region with 99.9\% light
absorption in comparison to the highly scattered values at 0.1\% stellar
absorption layer. This simple method could be also applied to estimate the
sensitivity of other parameters such as spectral features to $\beta$-factor, as
is discussed in \hyperref[sec:discussion]{Section}\;\ref{sec:discussion} .




\bibliographystyle{aasjournal}
\bibliography{Zotero}{}



\end{document}